\documentclass[aps,pra,reprint,twocolumn,showpacs,showkeys,longbibliography]{revtex4-1}
\usepackage{amsmath,amssymb,amstext}
\usepackage[usenames,dvipsnames]{color}
\usepackage{graphicx}
\usepackage{bm,bbold,braket}
\usepackage{natbib}
\usepackage{txfonts, comment,stmaryrd}
\usepackage{dcolumn}
\usepackage{color}
\usepackage{xcolor}
\colorlet{rn}{red}
\colorlet{an}{blue}
\usepackage[english]{babel}
\DeclareMathOperator{\sech}{sech}
\DeclareMathOperator{\csch}{csch}
\usepackage[colorlinks,bookmarks=false,citecolor=blue,linkcolor=red,urlcolor=blue]{hyperref}
\begin{document}

\title{Dynamics of a Pair of Overlapping Polar Bright Solitons in Spin-1 Bose-Einstein Condensates}
\author{Gautam Hegde}
\affiliation{Department of Physics, Indian Institute of Science Education and Research, Dr. Homi Bhabha Road, Pune 411 008, India}
\author{Sandra M Jose}
\affiliation{Department of Physics, Indian Institute of Science Education and Research, Dr. Homi Bhabha Road, Pune 411 008, India}
\author{Rejish Nath} 
\affiliation{Department of Physics, Indian Institute of Science Education and Research, Dr. Homi Bhabha Road, Pune 411 008, India}

\begin{abstract}
We analyze the dynamics of both population and spin densities, emerging from the spatial overlap between two distinct polar bright solitons in Spin-1 Spinor Condensates. The dynamics of overlapping solitons in scalar condensates exhibits soliton fusion, atomic switching from one soliton to another and repulsive dynamics depending on the extent of overlap and the relative phase between the solitons. The scalar case also helps us understand the dynamics of the vector solitons. In the spinor case, non-trivial dynamics emerge in spatial and spin degrees of freedom. In the absence of spin changing collisions, we observe Josephson-like oscillations in the population dynamics of each spin component. In this case, the population dynamics is independent of the relative phase, but the dynamics of the spin-density vector depends on it. The latter also witnesses the appearance of oscillating domain walls. The pair of overlapping polar solitons emerge as four ferromagnetic solitons irrespective of the initial phase difference for identical spin-dependent and spin-independent interaction strengths. But the dynamics of final solitons depends explicitly on the relative phase. Depending on the ratio of spin-dependent and spin-independent interaction strengths, a pair of oscillatons can also emerge as the final state. Then, increasing the extent of overlap may lead to the simultaneous formation of both a stationary ferromagnetic soltion and a pair of oscillatons depending on the relative phase.
\end{abstract}



\maketitle

\section{Introduction}
Because of the spin-dependent interactions, spinor condensates are ideal for exploring coherent spin-mixing dynamics \cite{kaw12,cha05, kro06, bla07,liu09, klem09, eto18, che19, evr21}. The resulting oscillations in the populations of the Zeeman states can be engineered by varying the initial populations and the relative phases \cite{hpu99}. In a spin-1 condensate, the oscillatory dynamics arises because of the collisional interconversion of two atoms in the $m=0$ state and one atom each in $m=1$ and $m=-1$ states. Such a spin-changing process preserves the net magnetization. Further, high controllability over the spin-dynamics can be accessed via external magnetic or microwave fields utilizing linear and quadratic Zeeman effects.

Spinor condensates also provide an opportunity to study vector solitons \cite{ber18, cha20, far20, lan20, liu20, cha21, men22} in quasi-one-dimensional (Q1D) BECs, including the bright solitons \cite{ied04, ied04b,llu05, sza11}. Vector solitons are self-trapped wave packets with multiple components. In spin-1 condensates, the bright solitons are classified into polar and ferromagnetic ones based on the spin state and the expectation value of the time-reversal operator \cite{sza11}. Collisional properties of polar-polar, polar-ferromagnetic and ferromagnetic-ferromagnetic solitons have been studied in the past \cite{ied04b, wan12} and that later lead to the discovery of an exotic soliton called the oscillaton \cite{sza11,sza10}. Oscillatons are solitons where the total density profile remains stationary while the populations of each $m$-components oscillate in time. It is also possible to observe oscillatory spin-dynamics in a single soliton \cite{zha07}, and in a pair of colliding solitons \cite{sza10}.

In this paper, we analyze the dynamics of a pair of overlapping polar bright solitons in spin-1 spinor condensates. Note that the scenario is different from the setups used to study soliton collisions. In the latter case, the solitons are initially placed very far apart and collide against each other with an initial velocity. The dynamics of overlapping optical solitons are previously studied both theoretically \cite{sax13,afa94,afa96} and experimentally \cite{sha91, sha92,ait91, ste99}, but similar studies of matter-wave solitons are lacking. Because of that, first, we look at the dynamics of overlapping bright solitons in scalar condensates. The latter helps us understand the unique features emerging from the vectorial nature of the spinor solitons. The dynamics critically depends on the initial phase difference and the extent of overlap between the solitons in the scalar case. It is expected that the relative phase plays a vital role since it decides the nature of force between solitons \cite{kha10}. Depending on the phase difference, the soliton fusion, atomic flow from one soliton to another and repulsive dynamics can occur. Similar scenarios are found in the dynamics of overlapping optical solitons in different non-linear media \cite{sax13,sha91,sha92,ait91}. In particular, the flow of atoms from one soliton to another mimics the phenomenon of optical switching, and we term it atomic switching for the matter-wave solitons.

The vectorial nature of spinor solitons leads to rich dynamics, both in population and spin densities. Besides the relative phase and extent of overlap between the solitons, the ratio between the spin-dependent and spin-independent interaction strengths also affects the dynamics. A simplified picture of the dynamics is attained using a rotated frame. In the absence of spin changing collisions, we observe Josephson-like oscillations in the population dynamics of each spin component. For atoms in each Zeeman state, the Josephson-like oscillations are explained using an effective potential created by the density of other components. In this case, the population dynamics are independent of the relative phase, but that of the spin-density vector depends on it. The dynamics of spin-density vector reveals the appearance of oscillating domain walls, depending on the relative phase. When spin-independent and spin-dependent interactions are identical, the collision of a pair of polar solitons results in four ferromagnetic solitons irrespective of the value of the initial relative phase. Among the four ferromagnetic solitons, each pair exhibits dynamics identical to that of scalar solitons that become more apparent in the new rotated frame. When the ratio of spin-dependent and spin-independent interactions is half, we see the formation of a pair of oscillatons. An additional stationary ferromagnetic soliton emerges if the extend of overlap is sufficiently large, depending on the relative phase.

The paper is structured as follows. In Sec. \ref{sc}, we discuss the dynamics of overlapping bright solitons in scalar condensates. In Sec. \ref{spin}, we discuss the dynamics of overlapping polar solitons in spin-1 condensates. In particular, the Sec. \ref{set} discusses the physical setup of spinor solitons, especially the governing Hamiltonian, the initial state, time-dependent coupled non-linear Schr\"odinger equations in the original and the rotated frame. In Sec.~\ref{dyn}, we classify the dynamics based on the ratio between the spin-dependent and spin-independent interaction strengths. In particular, the dynamics in the absence of spin-changing collisions is discussed in Sec.~\ref{g0}. The dynamics for identical spin-independent and spin-dependent interactions are discussed in Sec.~\ref{g1} and when the spin-dependent interaction strength is half of the spin-independent interaction is discussed in Sec.~\ref{gh}. Finally, we summarize in Sec.~\ref{sum}.

\section{Scalar condensates}
\label{sc}
\begin{figure}
\centering
\includegraphics[width= .9\columnwidth]{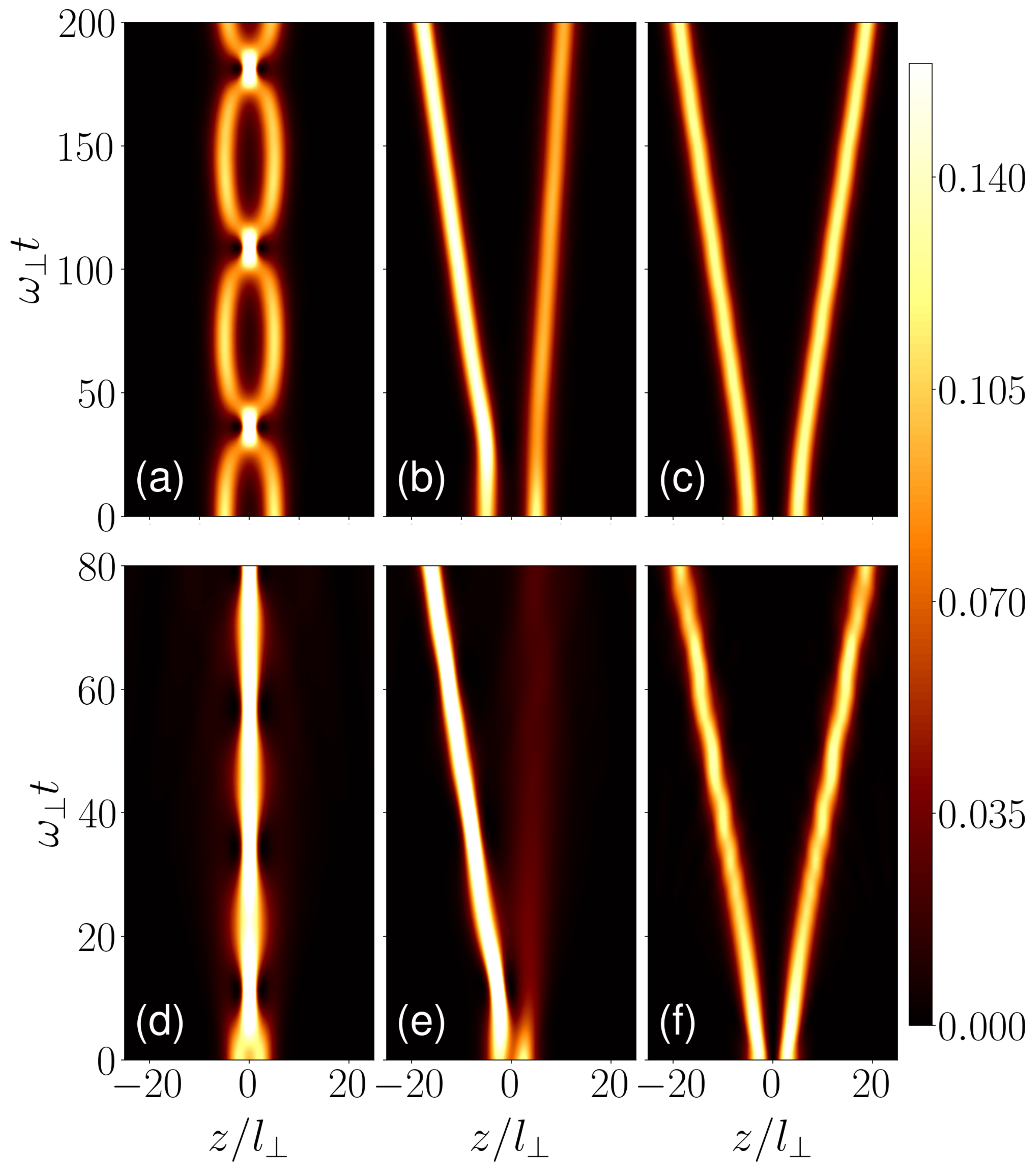}
\caption{\small{Dynamics of two overlapping identical bright solitons in scalar condensates for different relative phase between them for $g/(2\pi l_\perp^2)=-2$. The separation $\Delta=10l_\perp$ for (a)-(c) and $\Delta=5l_\perp$ for (d)-(f). (a) and (d) for $\phi=0$, and in (a) the overlap leads to perpetual oscillating dynamics between the soliton fusion and the initial configuration. In (d) the system did not reverse back completely to the initial configuration. (b) and (e) for $\phi=\pi/2$, an initial atomic flow along the phase gradient leads to asymmetric final solitons. (c) and (f) For $\phi=\pi$, the solitons repel each other, but identical in their size and shape. }}
\label{fig:1} 
\end{figure}

In the following, we analyze the dynamics of two identical, overlapping Q1D bright solitons in scalar condensates for different initial relative phases and overlaps. The analytic form of the initial two-soliton wavefunction is
\begin{equation}
  \psi(z,t=0)=A\left[\sech\left(k\left(z-\Delta/2\right)\right)+e^{i\phi}\sech\left(k\left(z+\Delta/2\right)\right)\right],
  \label{i1}
\end{equation}
where $\phi$ is the relative phase, $k$ is the wavenumber, $\Delta$ is the initial separation between the solitons, which controls the extent of overlap and the normalization constant is $$A=\frac{1}{2}\sqrt{\frac{k}{1+k\Delta\cos\phi\csch k\Delta}}.$$ For large values of $\Delta$, the solitons do not overlap, and they remain at rest, maintaining their size and shape over time. Interestingly, a tiny spatial overlap between the solitons can trigger non-trivial dynamics, depending critically on the phase difference. The nature of soliton interactions, whether being attractive or repulsive, also depends on their relative phase. The role of relative phase on the collisional dynamics of matter-wave solitons has been a subject of study both experimentally \cite{ngu14} and theoretically \cite{bil11}. Here, we analyze the dynamics mainly for three different relative phases: $\phi=0$, $\phi=\pi/2$ rad and $\phi=\pi$ rad, which captures the different dynamical scenarios. For the scalar condensates, the dynamics is governed by the Q1D non-linear Schr\"odinger equation,
\begin{equation}
i\hbar\frac{\partial}{\partial t}\psi(z, t)=\left[-\frac{\hbar^2}{2M}\frac{\partial^2}{\partial z^2}+\frac{g}{2\pi l_{\perp^2}}|\psi(z,t)|^2\right]\psi(z,t),
\end{equation}
where $g=4\pi\hbar^2 a_sN/M$ with the $s$-wave scattering length $a_s<0$, $N$ is the total number of particles and $l_{\perp}=\sqrt{\hbar/m\omega_{\perp}}$ is the width of the transverse harmonic confinement of frequency $\omega_{\perp}$. The wavenumber $k$ in Eq.~(\ref{i1}) depends on the interaction strength via $k=|g|/(2\pi\hbar\omega_\perp l_\perp^4)$.

\begin{figure}
\centering
\includegraphics[width= 1.\columnwidth]{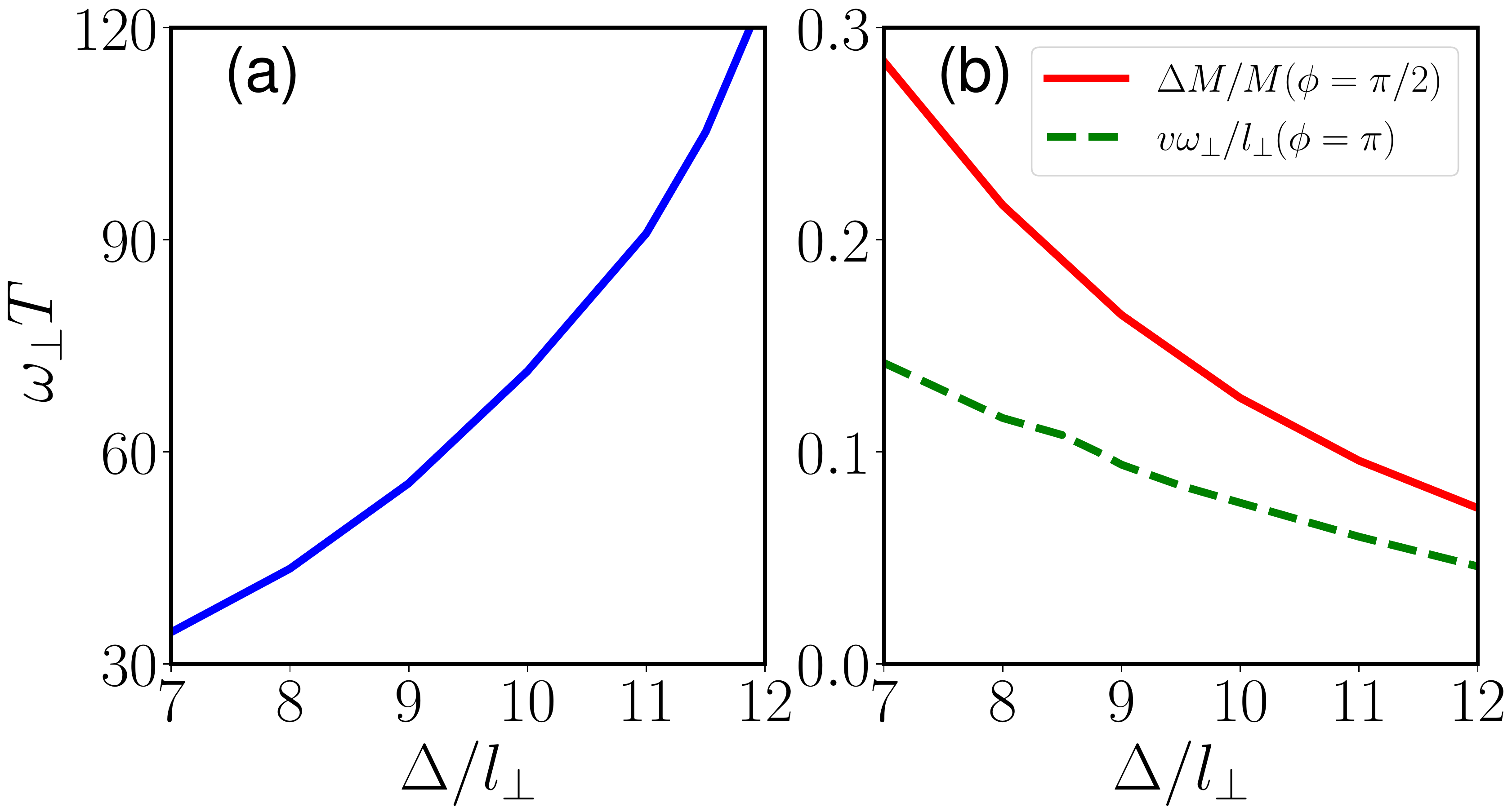}
\caption{\small{The soliton properties as a function of $\Delta$ for different initial relative phase for the scalar case. (a) The time period of oscillations between soliton fusion and the initial configuration for $\phi=0$. In (b), the solid line depicts the mass difference $\Delta M/M$ between the final solitons for $\phi=\pi/2$ rad and the dashed line shows the speed of the final symmetric solitons for $\phi=\pi$ rad. For both figures, $g/(2\pi l_\perp^2)=-2$. }}
\label{fig:2} 
\end{figure}

When $\phi=0$, the overlap leads to the attractive interaction between the solitons, making them fuse into a single soliton. Not being in its lowest energy state, the fused soliton disentangles back into the initial two-soliton configuration if the overlap is sufficiently small [see Fig.~\ref{fig:1}(a) for $\Delta=10 l_\perp$]. This process repeats periodically in time, leading to the perpetual oscillating dynamics between soliton fusion and the initial configuration as shown in Fig.~\ref{fig:1}(a). As the initial overlap between the solitons increases (decreasing $\Delta$), the period of oscillation gets shorter [see Fig.~\ref{fig:2}(a) for the time period $T$ vs $\Delta$]. Strikingly, oscillations in Fig.~\ref{fig:1}(a) arise in the absence of harmonic confinement, making it in high contrast with those exhibited by trapped solitons \cite{ngu14, bil11}. Beyond a certain overlap or smaller $\Delta$, the solitons become non-separable after the initial fusion as shown in Fig.~\ref{fig:1}(d) for $\Delta=5l_\perp$ and the final soliton exhibits breathing dynamics because of the extra energy it carries. These results are found to be similar to that of overlapping optical solitons in plasma \cite{sax13}.

As the initial relative phase between the solitons increases, they repel each other. For instance, when $\phi=\pi/2$ [see Fig.~\ref{fig:1}(b)], not only do the solitons repel, but also there is a flow of atoms along the direction of phase gradient, i.e., along $\partial\phi/\partial z$, at the initial stage of the dynamics. Similar to the atomic flow, an energy transfer among overlapping optical solitons has been observed, leading to optical switching \cite{sha91, sha92}. In a similar spirit, the atom transfer from one soliton to another can serve as a control knob to direct the motion of matter waves. Hence we call it atomic switching. In Fig.~\ref{fig:1}(b), the atomic flow takes place from the positive to negative $z$-direction. Such a transient atomic current makes the final solitons asymmetric in density, size and speed. The denser soliton moves faster. The speed of the lighter soliton arises from the recoil of the atoms flown out from the original soliton. To quantify the mass asymmetry, we introduce the mass difference between the left and right solitons, i.e., $\Delta M=M_l-M_r$, where $M_l=M\int_{-\infty}^0|\psi(z, t)|^2 dz$ and $M_r=M\int_0^{\infty}|\psi(z, t)|^2 dz$. In Fig.~\ref{fig:2}(b), the solid line shows $\Delta M$ obtained when the solitons are well separated after a sufficiently long time. As seen, $\Delta M$ increases with decreasing $\Delta$ because the larger the initial overlap more significant the exchange of particles. The atomic switching is evident in Figs.~\ref{fig:1}(b) and \ref{fig:1}(e), respectively for $\Delta=10l_\perp$ and $\Delta=5l_\perp$. 

\begin{figure}
\centering
\includegraphics[width= 1\columnwidth]{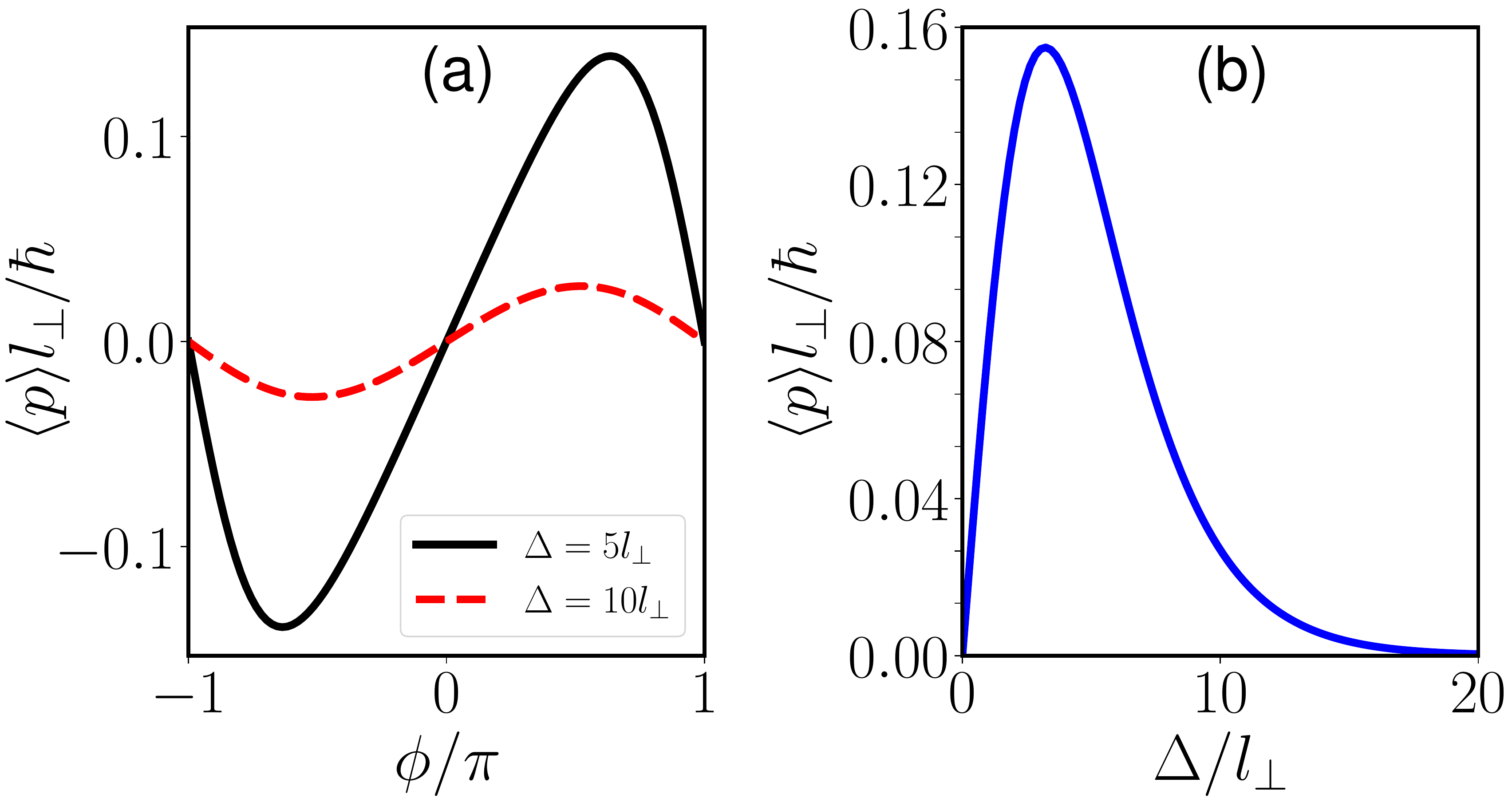}
\caption{\small{(a) The average momentum $\langle p\rangle$  as a function of $\phi$ for $\Delta=5 l_\perp$ (solid line) and $\Delta=10 l_\perp$ (dashed line) (b) $\langle p\rangle$ vs $\Delta$ for $\phi=\pi/2$ rad. $g/(2\pi l_\perp^3\hbar\omega_\perp)=-2$ for both (a) and (b).}}
\label{fig:3} 
\end{figure}

The above results imply that there is a net momentum in the system for $\phi=\pi/2$. To quantify that, we calculate the average momentum of the initial wave function in Eq.~(\ref{i1}) and is,
\begin{equation}
\langle p\rangle=4A^2\left(1-k\Delta\coth k\Delta\right)\csch k\Delta\sin\phi.
\label{pav}
\end{equation}
In Fig.~\ref{fig:3}(a), we show $\langle p\rangle$ as a function of $\phi$ for $\Delta=10l_\perp$ and $\Delta=5l_\perp$, which oscillates between positive and negative values as a function of $\phi$. The latter indicates that by tuning the relative phase, we can control the direction of the transient atomic current or the atomic switching. For $\phi=0$, $\langle p\rangle=0$ as expected, and for $\phi=\pi/2$ rad, there is a non-vanishing initial momentum leading to the asymmetric dynamics shown in Figs.~\ref{fig:1}(b) and \ref{fig:1}(e). For $\phi=\pi/2$ rad, $\langle p\rangle$ exhibits a non-monotonous behavior as a function of $\Delta$, exhibiting a local maximum, as shown in Fig.~\ref{fig:3}(b). At $\Delta=0$, the solitons completely overlap and remain at rest and thus $\langle p\rangle=0$. Also, as $\Delta\to\infty$, $\langle p\rangle$ approaches zero as both solitons become completely independent.

The solitons also repel each other for $\phi=\pi$ [see Figs.~\ref{fig:1}(c) and \ref{fig:1}(f)], but there is no transient atomic current due to the nodal point at $z=0$. Hence, the final solitons are identical, and $\Delta M=0$. The final solitons propagate with equal and opposite velocity, but the speed of each soliton depends on the separation $\Delta$. The final speed $v$ of the soliton vs $\Delta$ for $\phi=\pi$ rad is shown as a dashed line in Fig.~\ref{fig:2}(b). The more overlap the initial solitons have, the faster they move. Summarizing this section, we see that the initial phase difference and the extent of overlap critically affect the dynamics of overlapping bright solitons in scalar condensates. The most interesting feature is the local and transient atomic current in attractive overlapping condensates due to an effective repulsion arising from the phase difference between the solitons. This scenario is identical to optical switching, offering the possibility of engineering matter-wave transport via controlling the relative phase.

\section{Spin-1 Spinor condensates}
\label{spin}
\subsection{Setup, model, initial state and rotated frame}
\label{set}
The system we consider are partially overlapping spin-1 condensates, they are described by the Hamiltonian,
\begin{equation}
\hat H=\int dz\left[-\frac{\hbar^2}{2M}\sum_m\hat\psi_m^{\dagger}\frac{d^2}{dz^2}\hat\psi_m+\frac{1}{2} \bar c_0:\hat n^2:+\frac{1}{2}\bar c_1:\hat {\bm F}^2:\right],
\label{h1}
\end{equation}
where $M$ is the mass of a boson, $\bar c_{0, 1}=c_{0, 1}/2\pi l_{\perp}^2$ are the interaction parameters, $\hat\psi_m$ is the field operator which annihilates a boson of the $m$th Zeeman state, $\hat n(z)=\sum_{m=-f}^f\hat\psi_m^{\dagger}(z)\hat\psi_m(z)$ is the total density operator. The components of spin density operator are
\begin{equation}
\hat F_{\nu\in x, y, z}(z)=\sum_{m, m'}\left({\rm f}_\nu\right)_{mm'}\hat\psi_m^{\dagger}(z)\hat\psi_{m'}(z),
\end{equation}
with $\rm{f}_\nu$ being the $\nu$th component of the spin-1 matrices. The symbol : : denotes the normal ordering that places annihilation operators to the right of the creation operators. The spin-independent and spin-dependent interaction parameters are $c_0=(g_0+2g_2)/3$ and $c_1=(g_2-g_0)/3$, respectively with $g_{\mathcal F}=4\pi\hbar^2a_{\mathcal F}N/m$ related to the scattering length $a_{\mathcal F=0, \ 2}$ of the total spin-${\mathcal F}$ channel. $N$ is the total number of atoms. Since we are interested in bright solitons, we always keep $c_0<0$.

Within the mean-field theory, the dynamics of the system is described by the quasi-one-dimensional (Q1D) Gross-Pitaevskii equations (GPEs) \cite{ied04,ied04b,sza10}
\begin{eqnarray}
\label{m1}
i\hbar\frac{\partial \psi_{1}}{\partial t}=\left[-\frac{\hbar^2}{2M}\frac{\partial^2}{\partial z^2}+\bar c_0n+\bar c_1 F_z\right]\psi_1+\frac{c_1}{\sqrt{2}} F_-\psi_0, \\
\label{m2}
i\hbar\frac{\partial \psi_{0}}{\partial t}=\left[-\frac{\hbar^2}{2M}\frac{\partial^2}{\partial z^2}+\bar c_0n\right]\psi_0+\frac{\bar c_1}{\sqrt{2}} F_+\psi_1+\frac{\bar c_1}{\sqrt{2}} F_-\psi_{-1}, \\
\label{m3}
i\hbar\frac{\partial \psi_{-1}}{\partial t}=\left[-\frac{\hbar^2}{2M}\frac{\partial^2}{\partial z^2}+\bar c_0n-\bar c_1F_z\right]\psi_{-1}+\frac{\bar c_1}{\sqrt{2}}F_+\psi_{0},
\end{eqnarray}
where  $n(z, t)=\sum_m|\psi_m(z, t)|^2$ is the total density, $F_\nu(z, t)=\sum_{m,m'}\psi^*_m(\rm{f}_\nu)_{mm'}\psi_{m'}$, and $F_{\pm}=F_x\pm i F_y$. We introduce $\gamma=-c_1/|c_0|$ as the ratio of spin-dependent and spin-independent interactions. The validity of Eqs.~(\ref{m1})-(\ref{m3}) requires that $\mu_{1D}\ll\hbar\omega_{\perp}$, where $\omega_\perp$ is the transverse confinement frequency and $\mu_{1D}$ is the chemical potential of the Q1D condensates. We solve Eqs.~(\ref{m1})-(\ref{m3}) numerically to analyze the dynamics \cite{sym16}. At $\gamma=1$, Eqs.~(\ref{m1})-(\ref{m3}) represent a completely integrable system and support $N$-soliton solutions including two-soliton ones \cite{ied04}.

\begin{figure}
\centering
\includegraphics[width= .75\columnwidth]{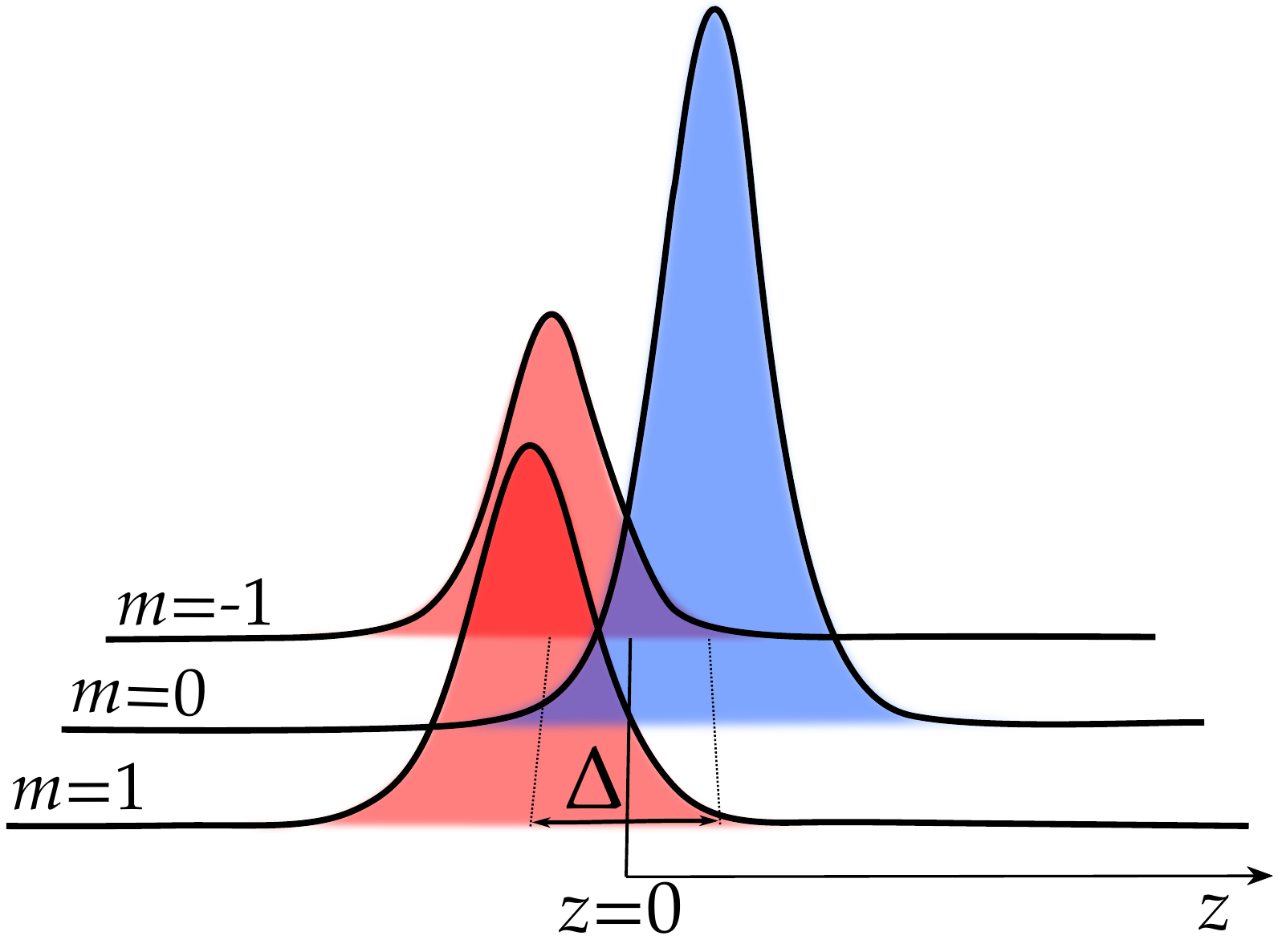}
\caption{\small{Schematic setup of overlapping bright polar solitons in Q1D spin-1 condensates, along $z$-axis. A polar soliton with population shared among $m=\pm 1$ states on the left (red-shaded) and a polar soliton with population solely at $m=0$ on the right side (blue-shaded). The separation $\Delta$ determines the extent of overlap between the two solitons.}}
\label{fig:0} 
\end{figure}

{\em Initial state}: The schematic diagram of the initial state is shown in Fig.~\ref{fig:0}, in which two distinct polar solitons overlap around $z=0$. The general solution of a static polar soliton is,
\begin{equation}
\psi(z)=\sqrt{\frac{k}{2}} \chi \sech kz,   
\end{equation}
where $k=|\bar c_0|/(4l_\perp)$ is the inverse-width or the wavenumber of the soliton wavepacket, and $\chi$ is the spin state, which takes the general form \cite{sza11,sza10},
\begin{equation}
\chi=e^{i\tau}\left(\begin{array}{c}-\frac{1}{\sqrt{2}} \sin \theta \\ \cos \theta \\ \frac{1}{\sqrt{2}} e^{i \phi} \sin \theta\end{array}\right),
\label{pv}
\end{equation}
where $\tau$ is a global phase. The initial wave function of a pair of overlapping polar solitons we consider is, 
\begin{equation}
    \psi(z)=A\left[\sech\left(k\left(z+\Delta/2\right)\right) \chi_l+\sech \left(k\left(z-\Delta/2\right)\right) \chi_r\right],
    \label{oss}
\end{equation}
where $A=\sqrt{k}/2$ is the normalization constant, and $\chi_r=(0, e^{i \phi_1}, 0)^T$ and $\chi_l=(1, 0, e^{i \phi_2})^T/\sqrt{2}$ are the spin states of the solitons in right and left of $z=0$, respectively. Note that, individually both polar solitons are degenerate ground state solutions of the time-independent GPEs for $\bar c_1>0$. Comparing to Eq.~(\ref{pv}), $\theta=\pi/2$ for the left soliton and that of the right one is $\theta=0$. The angles, $\phi_1$ and $\phi_2$ are the initial phases of the individual solitons and the extent of overlap between the solitons is again controlled by $\Delta$. For Eq.~(\ref{oss}), $\langle p\rangle=0$, independent of the value of $\Delta$ and $\phi_1$.

The spin density vector is a null vector for polar solitons \cite{ied04,ied04b,sza11}, but once they overlap, it may take a non-zero value in the overlapping region. For the initial state in Eq.~(\ref{oss}), the local spin density vector is,
\begin{equation}
\label{sdv}
{\bf F}(z, t=0)=\frac{4A^2\cos (\phi_1-\phi_2/2)}{\cosh 2kz+\cosh k\Delta}\left(\hat x\cos\frac{\phi_2}{2} +\hat y\sin\frac{\phi_2}{2}\right).
\end{equation}
The vector ${\bf F}(z, t=0)$ lies in the $xy$-plane and forms an angle $\phi_2$ with the $x$-axis. The orientation is same at every point along the $z$-axis, and depends only on $\phi_2$. The relative angle, $\phi_1-\phi_2/2$ determines the magnitude of ${\bf F}(z, t=0)$ and hence, the net magnetization. Integrating over $z$, we obtain the net spin density vector,
\begin{equation}
{\bf F}_T=4A^2\Delta\cos (\phi_1-\phi_2/2)\csch k\Delta \left(\hat x\cos\frac{\phi_2}{2} +\hat y\sin\frac{\phi_2}{2}\right),
\end{equation}
which is a conserved quantity. Without loss of generality, we set $\phi_2=0$, which fixes the direction of ${\bf F}_T$ along the $x$-axis. In the following we analyze the dynamics mainly for $\phi_1=0, \ \pi/2$ and $\pi$ rad. For $\phi_1=0$ and $\phi_1=\pi$ rad, the initial spin-density vector is along the positive and negative $x$-axis, respectively whereas for $\phi_1=\pi/2$ rad, ${\bf F}(z, t=0)$ vanishes. In the overlapping region, for $\gamma\neq 0$, spin-mixing dynamics takes place, in which two atoms in $m=0$ collisionally convert into one atom each in $m=1$ and $m=-1$, and vice versa. As shown below, the nature of dynamics can be classified based on the value of $\gamma$ and $\phi_1$.

\label{gl1}
\begin{figure}
\centering
\includegraphics[width= \columnwidth]{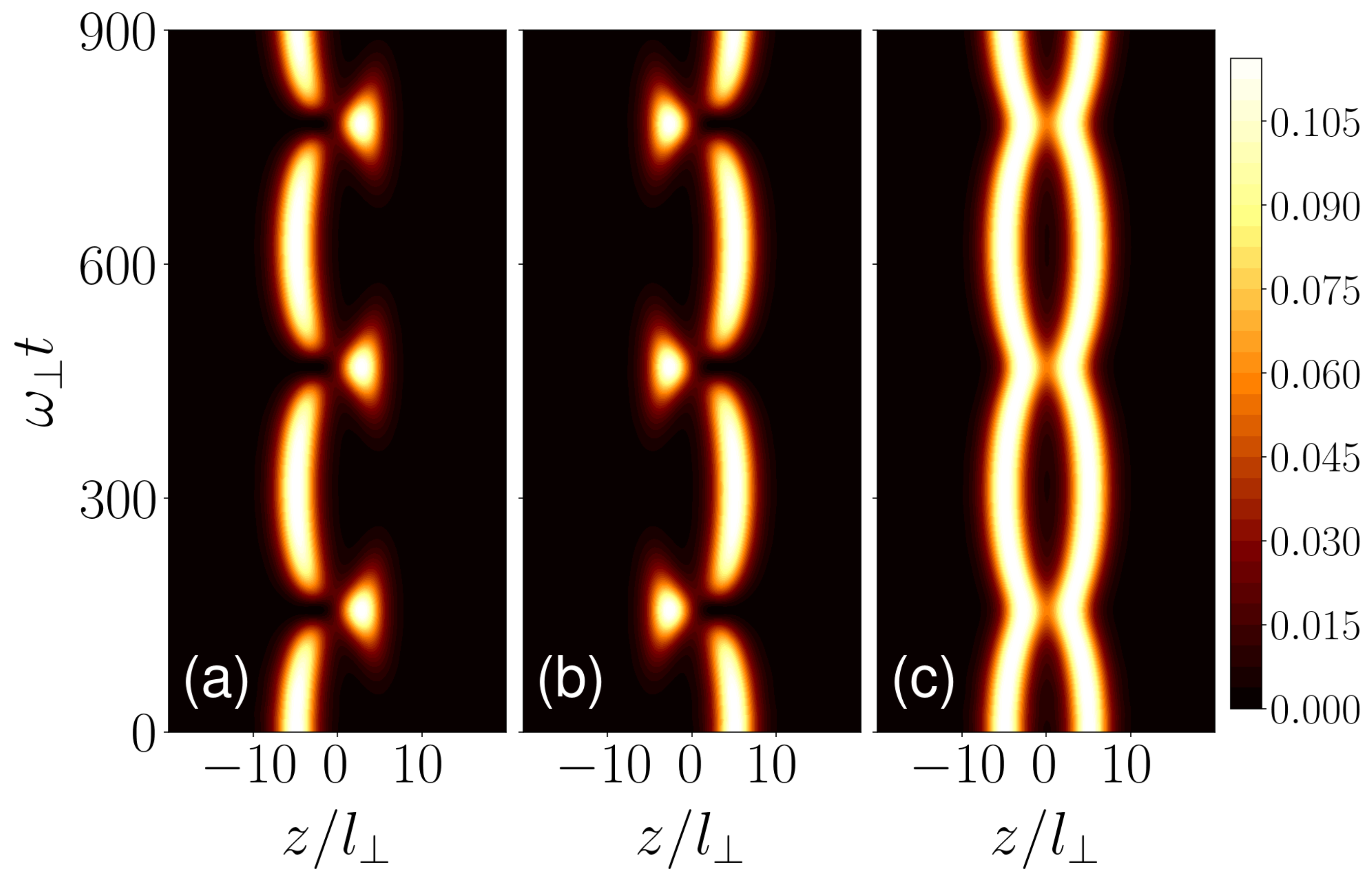}
\caption{\small{Dynamics of two overlapping polar bright solitons for $\gamma=0$, $\Delta=10l_\perp$ and $\bar c_0/\hbar\omega_\perp=-2$. (a)-(c) show the dynamics of $|\psi_1|^2+|\psi_{-1}|^2$, $|\psi_0|^2$ and the total density $n(z, t)$, respectively. The results are independent of the value of $\phi_1$.}}
\label{fig:4} 
\end{figure}
{\em Rotated frame}: Since ${\bf F}_T$ is a conserved quantity, we can move to a frame in which the quantization axis is parallel to the direction of ${\bf F}_T$ \cite{zha07}, which provides us with a simplified picture for the dynamics. Such a rotated frame is used in \cite{sza11,sza10} to describe oscillatons, which emerged out of the collision between a ferromagnetic and a polar soliton. In our case, the new frame is obtained by a rotation of $\pi/2$ about the $y$-axis, ($\chi'=e^{i\pi \rm{f}_y/2}\chi$) i.e., making the quantization axis along the $x$-axis. The relation between the new and old spinor components are
\begin{eqnarray}
\label{re1}
\psi_1'(z)=\frac{1}{2}\psi_1(z)+\frac{1}{\sqrt{2}}\psi_0(z)+\frac{1}{2}\psi_{-1}(z) \\
\label{re2}
\psi_0'(z)=-\frac{1}{\sqrt{2}}\psi_1(z)+\frac{1}{\sqrt{2}}\psi_{-1}(z) \\
\label{re3}
\psi_{-1}'(z)=\left(\frac{1}{2}\psi_1(z)-\frac{1}{\sqrt{2}}\psi_0(z)+\frac{1}{2}\psi_{-1}(z)\right).
\end{eqnarray}
Note that for the setup we consider $\psi_1$ and $\psi_{-1}$ are identical at any instant, and hence $\psi_0'(z)$ completely vanishes. Hence, in the new frame, the system effectively reduces to a two-component condensate \cite{lee16}. The corresponding GPEs are,
\begin{eqnarray}
\label{efe}
i\hbar\frac{\partial \psi'_{\pm 1}}{\partial t}&=&\left[-\frac{\hbar^{2}}{2 M}\frac{\partial^2}{\partial z^2}+\bar c_+ |\psi_{\pm 1}^\prime|^2 + \bar c_-|\psi_{\mp 1}^\prime|^2 \right] \psi_{\pm1}^\prime, 
\end{eqnarray}
where $\bar c_\pm=\bar c_0 \pm \bar c_{1}$. The exchange coupling between the two components $\psi'_{\pm 1}$ is provided by $\bar c_0-\bar c_1$. When $\gamma=1$, $\bar c_0-\bar c_1$ vanishes, indicating that in the rotated frame, the system effectively reduces to two independent scalar condensates described by the wave functions $\psi'_{1}$ and $\psi'_{-1}$. Note that $\psi_1'$ and $\psi_{-1}'$, in general do not exhibit identical dynamics. We use the rotated frame to gain insights into the dynamics wherever possible.
\subsection{Dynamics}
\label{dyn}
A comprehensive study of the dynamics of overlapping polar solitons as a function of $\Delta$, $\gamma$ and $\phi_1$ is a tedious task. We restrict the analysis to $\phi_1=0, \ \pi/2, \ \pi$ rad, $\Delta=5l_\perp, \  10l_\perp$ and $0\leq\gamma\leq 1$ and in particular, (i) $\gamma=0$, (ii) $\gamma=1$ and (iii) $\gamma=0.5$, which capture the most exciting scenarios. 
\subsubsection{$\gamma=0$}
\label{g0}
First, we discuss the case for which there are no spin changing collisions, i.e., $\gamma=0$. In this case, the total population in each Zeeman component remains constant, and the population dynamics is found to be independent of $\phi_1$. We also found that changing $\Delta$ does not affect the dynamics qualitatively. The dynamics for $\gamma=0$ and $\Delta=10l_\perp$ is shown in Fig.~\ref{fig:4}. In particular, Figs.~\ref{fig:4}(a)-\ref{fig:4}(c) show the dynamics of $|\psi_1|^2+|\psi_{-1}|^2$, $|\psi_0|^2$ and the total density $n(z, t)$, respectively. Effectively, we observe Josephson-like oscillations of populations in each component [see Figs.~\ref{fig:4}(a) and \ref{fig:4}(b)] between the two regions (left and right of $z=0$) where the solitons are initially placed. The oscillations are non-sinusoidal. The density oscillations may give a false impression that spin-changing collisions are taking place. To demonstrate the Josephson-like oscillations, for each Zeeman state, we introduce the population imbalance between the left and right sides of $z=0$, i.e., 
\begin{equation}
Z_m(t)=\frac{1}{N_m}\left(\int_{-\infty}^0dz|\psi_m|^2-\int_0^{\infty}dz|\psi_m|^2\right),
\end{equation}
where $N_m$ is the total number of atoms in $m$th state, which is a constant.
\begin{figure}
\centering
\includegraphics[width= \columnwidth]{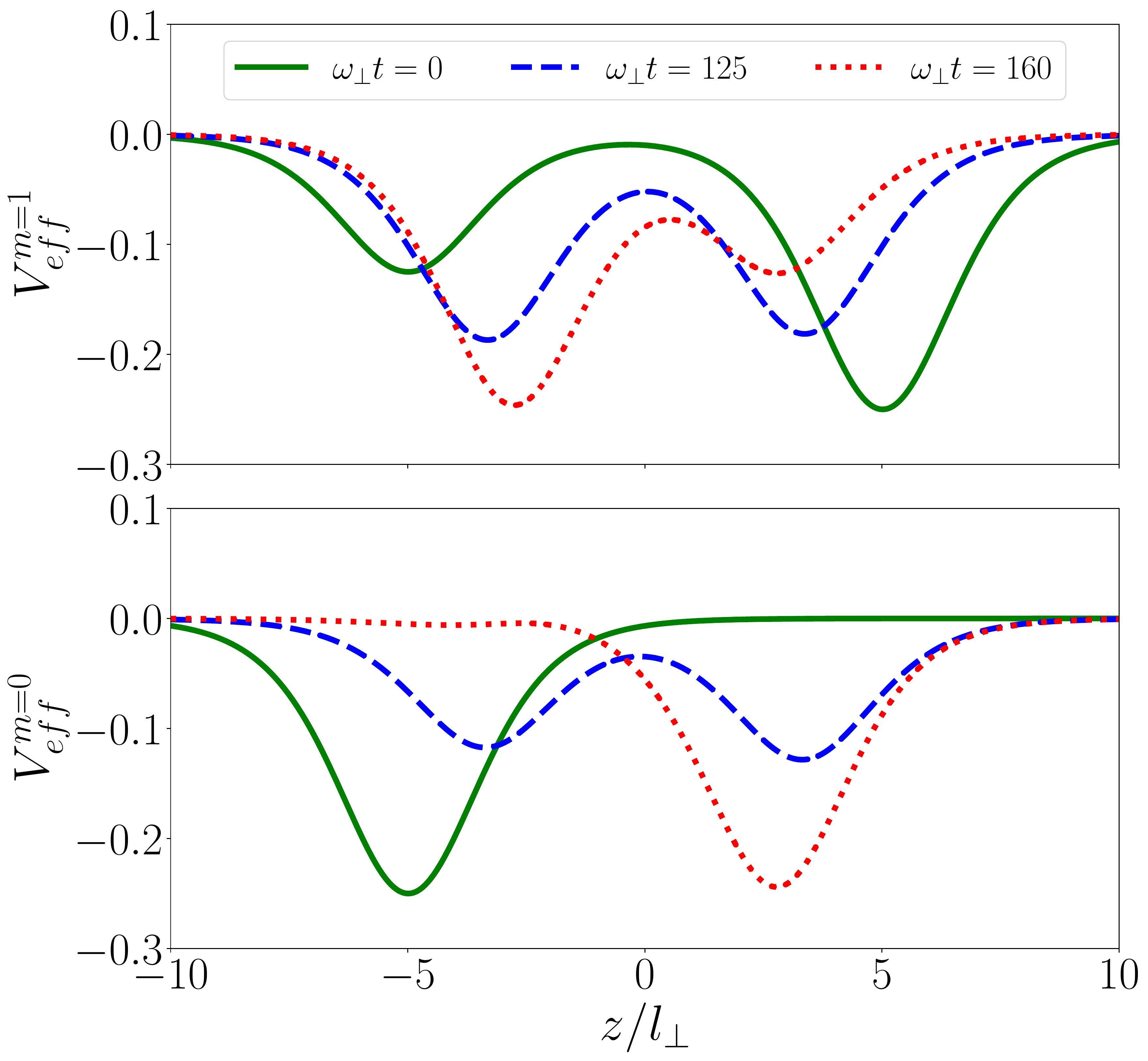}
\caption{\small{Effective potential $V_{eff}^m$ generated for $m$th component by other components for $\gamma=0$, $\Delta=10l_\perp$ and $\bar c_0/\hbar\omega_\perp=-2$. (a)$V_{eff}^{m=1}$ exhibits a double well potential with its local minima and separation between them varying in time. (a)$V_{eff}^{m=0}$ oscillates in time between a single minimum and double minima. The results are independent of the value of $\phi_1$}}
\label{fig:5} 
\end{figure}

The density oscillations of each component can be intuitively understand via an effective potential created by other components.  First consider the population dynamics of $m=1$ state, which is shown in Fig.~\ref{fig:4}(a) (identical for $m=-1$ state also). Since most of the population in $m=1$ is initially at the left side of $z=0$, we have $Z_1(t=0)\sim 1$. For $\bar c_0<0$, the terms $\bar c_0|\psi_{-1}|^2$ and  $\bar c_0|\psi_0|^2$ in Eq.~(\ref{m1}) form a double well potential for atoms occupying $m=1$ state, see Fig.~\ref{fig:5}(a) for $V_{eff}^{m=1}(z,t)=\bar c_0(|\psi_0|^2+|\psi_{-1}|^2)$ at different instants. A non-zero fraction of $m=1$ atoms in the right region triggers the tunnelling from left to right, leading eventually to oscillatory dynamics shown in Fig.~\ref{fig:4}(a). At $t=0$, $V_{eff}^{m=1}$ is asymmetric in $z$ since $m=0$ population on the right side is twice that of $m=-1$ state on the left. As time evolves, $V_{eff}^{m=1}$ gets modified, in particular, the two minima gets closer to each other [see dashed and dotted lines in Fig.~\ref{fig:5}(a)], which amplifies the tunnelling rate. Eventually the potential gets inverted but with a shorter separation between the minima, as shown by dotted line in Fig.~\ref{fig:5}(a). The potential gets inverted because of the swapping of population in $m=0$ ($m=-1$) from left (right) to right (left). By this time, the majority of population in $m=1$ state has already tunnelled to the right. At this points, the reverse dynamic happens, and the system recover to the initial density configuration. The whole processes then repeats periodically in time.

\begin{figure}
\centering
\includegraphics[width= \columnwidth]{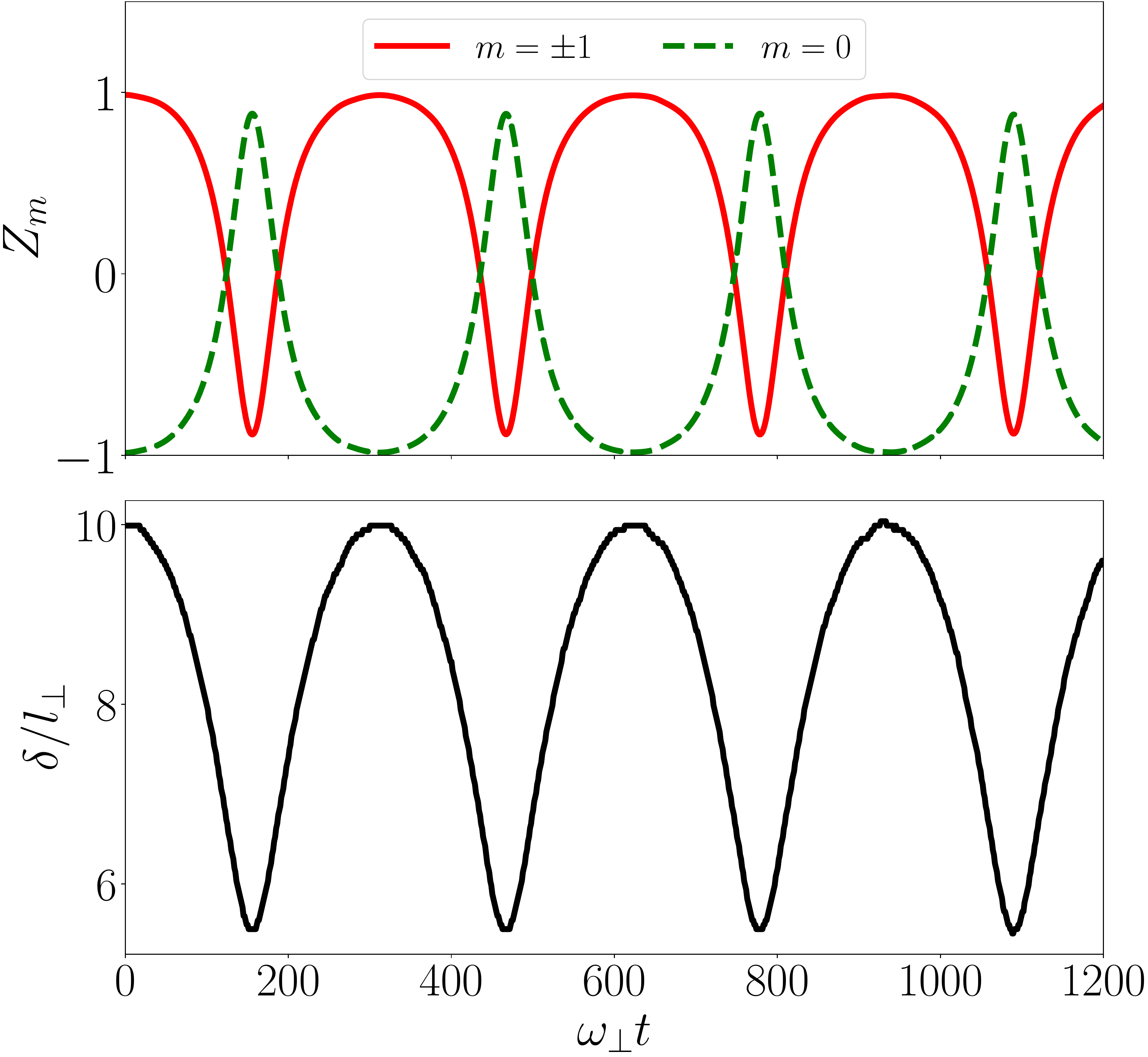}
\caption{\small{Dynamics of (a) the population imbalance $Z_m$ and (b) the separation $\delta(t)$ between the two peaks in the total density $\gamma=0$, $\Delta=10l_\perp$ and $\bar c_0/\hbar\omega_\perp=-2$. The results are independent of the value of $\phi_1$.}}
\label{fig:6} 
\end{figure}

An identical picture also holds for the population dynamics of the $m=-1$ state. In contrast, the population in $m=0$, initially placed in the right side, experiences a different dynamical potential [$V^{m=0}_{eff}(z, t)=\bar c_0(|\psi_1|^2+|\psi_{-1}|^2)$] from the populations of $m=\pm 1$, as shown in Fig.~\ref{fig:5}(b). As time evolves, the potential minimum, initially on the left (solid line), moves to the right via a transient double-well potential and then reverts. The $m=0$ atoms always move towards the potential minimum in the opposite region leading to the oscillatory dynamics shown in Fig.~\ref{fig:4}(b). Also, note that $m=\pm 1$ populations remain immiscible with the $m=0$ component for the entire time, except in the overlapping region. The total density is characterized by an out-of-phase oscillatory dynamics of the two peaks, see Fig.~\ref{fig:4}(c).

In Fig.~\ref{fig:6}(a), we show the dynamics of population imbalance $Z_m$ for the same dynamics shown in Fig.~\ref{fig:4}. Figure~\ref{fig:6}(b) shows the separation ($\delta$) between the two peaks in the total density or equivalently the distance between the two minima in $V_{eff}^{m=1}(z,t)$ [see Fig.~\ref{fig:4}(c)]. At $t=0$, $\delta=\Delta$ and it varies periodically in time. The population dynamics get faster as $\delta$ decreases and vice versa. It is seen in Fig.~\ref{fig:6}(a) that the population of each component in the initial region never vanishes. Thus, there are always some atoms in $m=\pm 1$ states on the left region and $m=0$ on the right region, which helps the system periodically recover to initial densities. 

\label{gl1}
\begin{figure}
\centering
\includegraphics[width= \columnwidth]{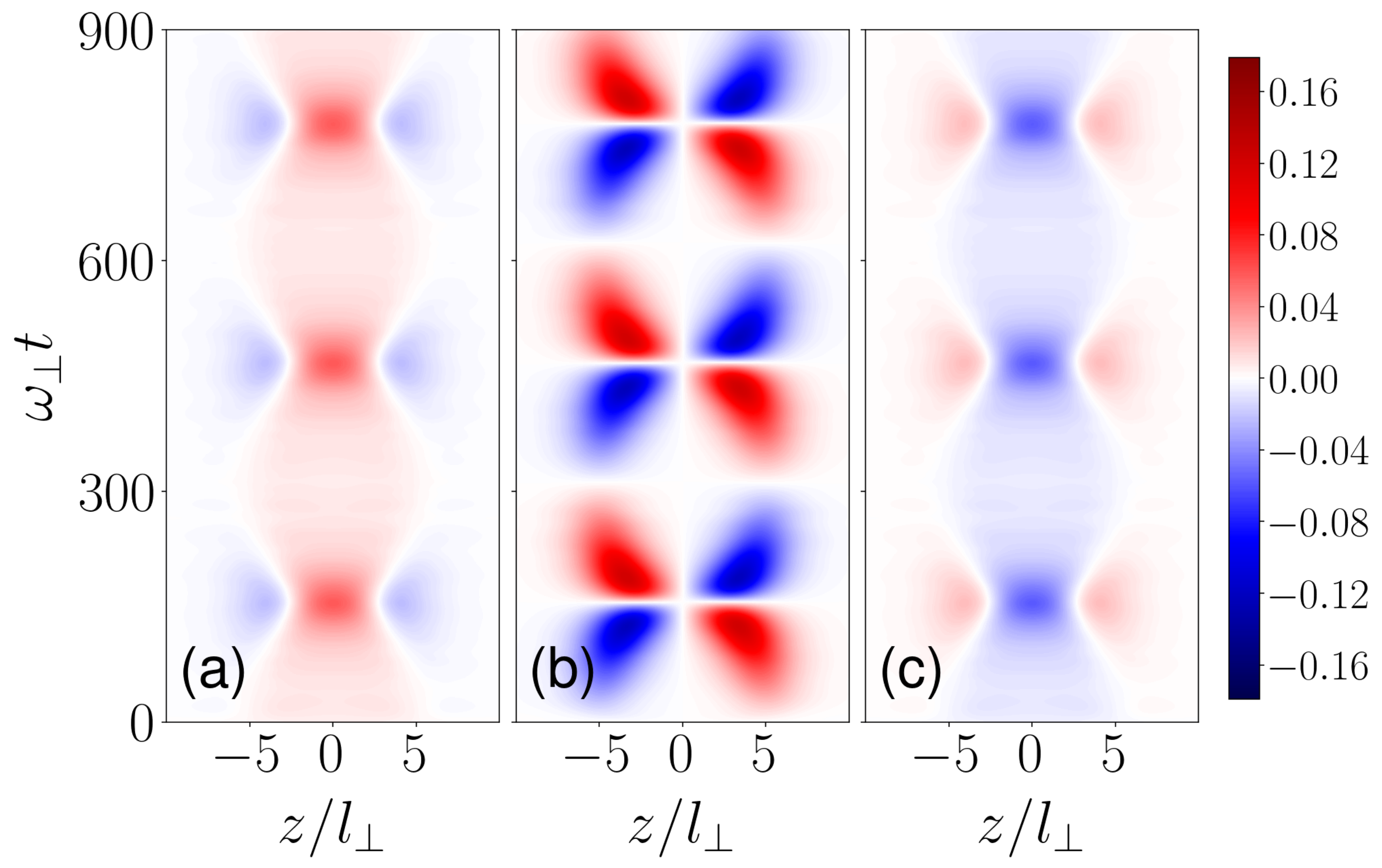}
\caption{\small{Dynamics of spin-density $F_x(z, t)$ for (a) $\phi_1=0$, (b) $\phi_1=\pi/2$ rad, and (c) $\phi_1=\pi$ rad. Other parameters are $\gamma=0$, $\Delta=10l_\perp$ and $\bar c_0/\hbar\omega_\perp=-2$.}}
\label{fig:7} 
\end{figure}
Even though the population dynamics shown in Fig.~\ref{fig:4} is independent of $\phi_1$, it affects the spin density vector ${\bf F}(z, t=0)= F_x(z,t)\hat x$ via Eq.~(\ref{sdv}) and its dynamics. In Figs.~\ref{fig:7}(a)-\ref{fig:7}(c), we show the dynamics of $F_x(z, t)$ for $\phi_1=0$, $\phi_1=\pi/2$ rad, and $\phi_1=\pi$ rad, respectively. Since $F_x(z,t)\propto (\psi_1+\psi_{-1})\psi_0^*+(\psi_1^*+\psi_{-1}^*)\psi_0$, the spin density vector is significant only in the overlapping regions. For $\phi_1=0$, $F_T=\int_{-\infty}^{\infty}F_x(z, t) dz$ is positive, i.e., the effective spin is pointing along the positive $x$-direction. At $t=0$, the spin density vector is along the $x$-axis at every point in the overlapping region. As time progresses, we observe the formation of domains with positive and negative $F_x$. In particular, regions of negative values of $F_x$ emerge from both the edges of the overlapping region, separated by a region of positive $F_x$ [see Fig.~\ref{fig:7}(a)]. The domain walls, which separate the regions of positive and negative $F_x$, move towards the centre, squeezing the region of positive $F_x$ in the middle. The latter results in an increase in spin density at the centre. Because of the conservation of ${\bf F}_T$ and energy, they cannot reach beyond a separation. Eventually, the domain walls move back to the edges, and their position oscillates in time. For $\phi_1=\pi$ rad, the dynamics of $F(z, t)$ is identical to that of $\phi_1=0$ except that positive and negative regions are switched as shown in Fig.~\ref{fig:7}(c).
 
The dynamics of $F(z, t)$ for $\phi_1=\pi/2$ rad is drastically different from that of $\phi_1=0$ and $\phi_1=\pi$ rad, see Fig.~\ref{fig:7}(b). For $\phi_1=\pi/2$ rad, and $F(z, 0)$ vanishes and hence $F_T$. As time progresses, we see simultaneous growth of equal regions of positive and negative $F_x$. The growth happens from the edges of the overlapping region, separated by a region of vanishing $F_x$. As time evolves, the size of the central region of $F_x=0$ shrinks and the regions of $F_x\neq 0$ grow. After some time, surprisingly, $F_x(z,t)$ vanishes quite rapidly and then reemerges with opposite polarity. Then, the central region with $F_x=0$ grows until the regions of $F_x\neq 0$ on either sides diminish. The whole process repeats periodically in time and leads to the butterfly pattern in space-time, as seen in Fig.~\ref{fig:7}(b). Thus, we have a unique scenario of engineering the spin dynamics in spinor condensates leaving the population dynamics unaffected, by tuning the relative phase $\phi_1$, in the absence of spin-independent interactions.
  
\subsubsection{$\gamma=1$}
\label{g1}
\begin{figure}
\centering
\includegraphics[width= \columnwidth]{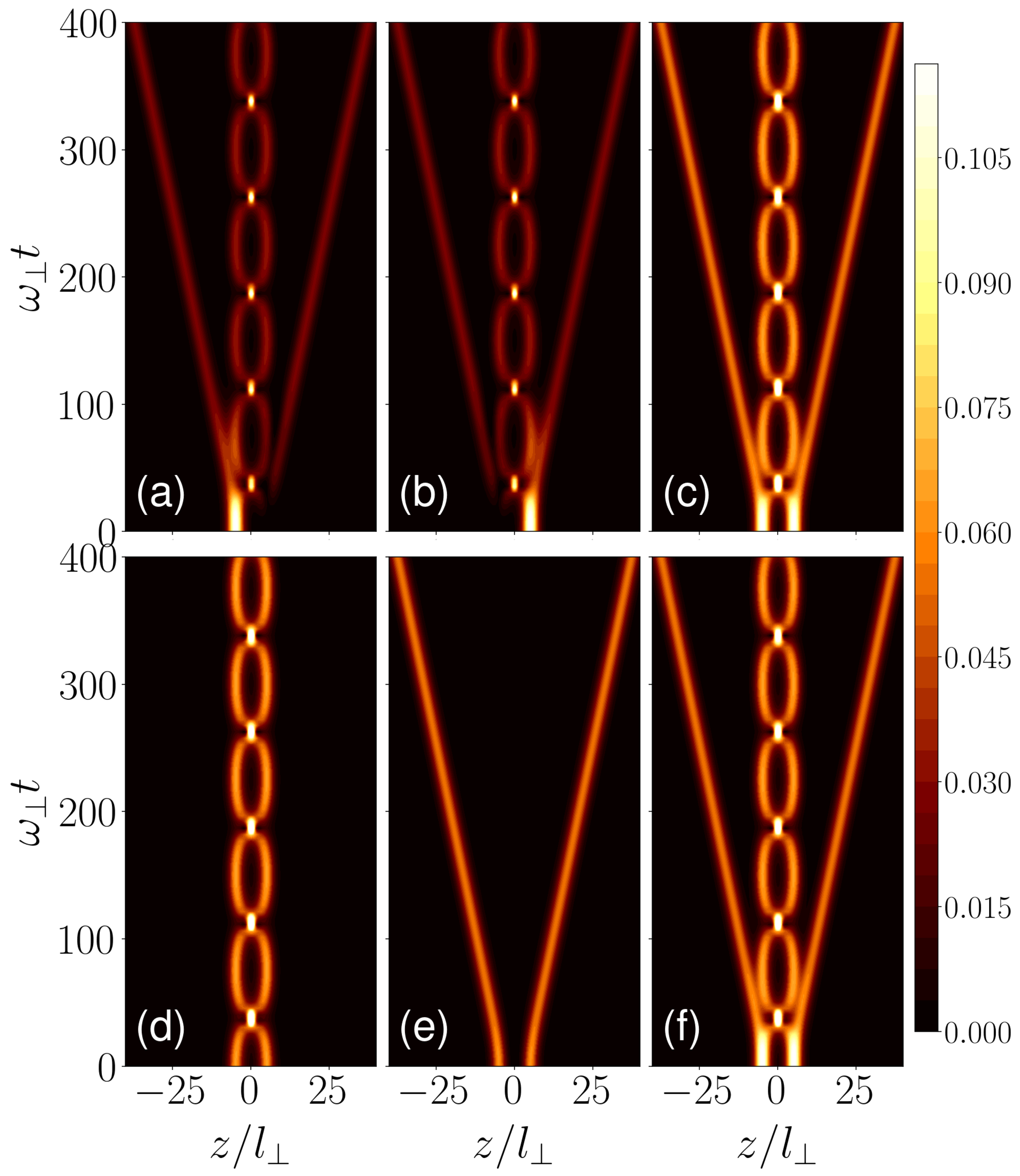}
\caption{\small{Dynamics of two overlapping polar bright solitons for $\gamma=1$, $\Delta=10l_\perp$, $\bar c_0/\hbar\omega_\perp=-2$, and $\phi_1=0$. (a)-(c) show the dynamics of $|\psi_1|^2+|\psi_{-1}|^2$, $|\psi_0|^2$ and the total density $n(z, t)$, respectively. (d)-(f) show the dynamics of $|\psi'_1|^2$, $|\psi'_{-1}|^2$ and the total density $n'(z, t)=|\psi'_1|^2+|\psi'_{-1}|^2$, respectively. Note that $n(z,t)=n'(z,t)$.}}
\label{fig:8} 
\end{figure}

\begin{figure}
\centering
\includegraphics[width= \columnwidth]{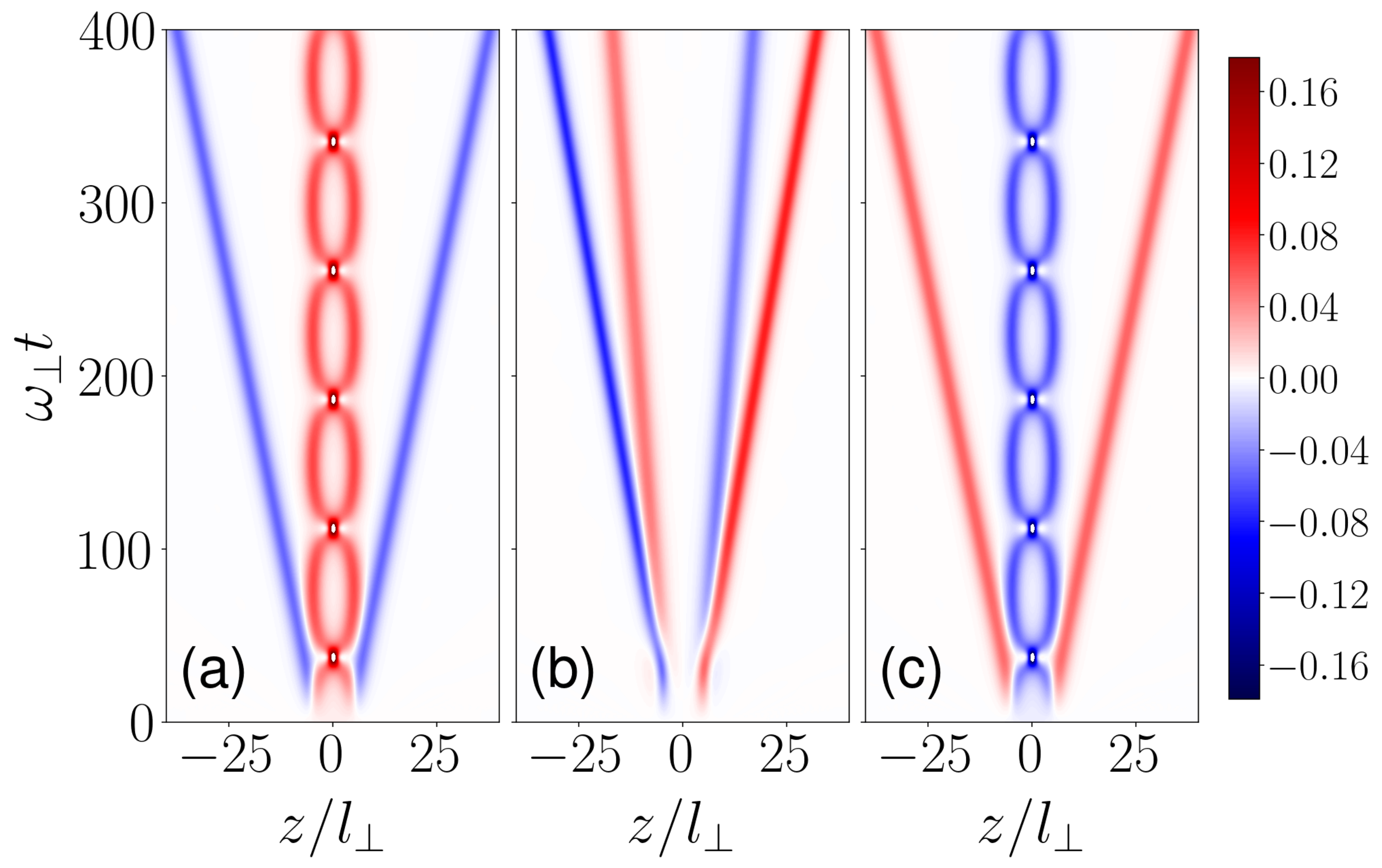}
\caption{\small{Dynamics of spin-density $F_x(z, t)$ for (a) $\phi_1=0$, (b) $\phi_1=\pi/2$ rad, and (c) $\phi_1=\pi$ rad. Other parameters are $\gamma=1$ and $\Delta=10l_\perp$.}}
\label{fig:9} 
\end{figure}

The presence of spin-changing collisions affects the dynamics drastically. For $\gamma=1$, spin-dependent and spin-independent interactions are of equal strength. In Figs.~\ref{fig:8}(a)-\ref{fig:8}(c), we show the dynamics of $|\psi_1|^2+|\psi_{-1}|^2$, $|\psi_0|^2$ and the total density $n(z, t)$, respectively for $\gamma=1$, $\Delta=10 l_\perp$ and $\phi_1=0$. The first visible effect of spin-changing collision is that all Zeeman components become spatially miscible, leading to identical density patterns for all Zeeman components at longer times. After a sufficiently long time, the initial, overlapping polar solitons converted into four solitons via spin-changing collisions, two each on either side of $z=0$. Each soliton is characterized by a population ratio of 1 : 2 : 1 among the Zeeman states ($m=-1$, $m=0$, $m=1$) with a spin state $\chi=(1, \pm\sqrt{2}, 1)^T/2$, which is a ferromagnetic soliton \cite{sza11,sza10}. The spin density vector shown in Fig.~\ref{fig:9}(a) confirms that the final solitons are indeed ferromagnetic but it possesses opposite directions for the inner and outer solitons. The inner solitons have $F_x(z, t)>0$, i.e., the polarization axis is along the positive $x$-axis, whereas the outer ones have the net spin vector along the negative $x$-axis. Thus, we have a scenario where two overlapping polar solitons are dynamically converted into four ferromagnetic solitons.

In the rotated frame [Eqs.~(\ref{re1})-(\ref{re3})], the scenario becomes relatively simple and also provide additional details on the dynamics. Recall that for $\gamma=1$, the wavefunctions $\psi_1'(z)$ and $\psi'_{-1}(z)$ are decoupled, and effectively we have two independent scalar condensates. At $t=0$, the initial states with $\phi_1=0$ are, 
\begin{eqnarray}
\label{re01}
\psi_{\pm 1}'(z)&=&\frac{\sqrt{k}}{2}\left[\frac{\sech\left(k\left(z-\Delta/2\right)\right)}{\sqrt{2}}\pm\frac{\sech\left(k\left(z+\Delta/2\right)\right)}{\sqrt{2}}\right].
\end{eqnarray}
As seen in Eq.~(\ref{re01}), $\psi'_{1}$ and $\psi'_{-1}$ represent two-soliton states, identical to Eq.~(\ref{i1}), with relative phase $\phi=0$ and $\phi=\pi$ rad, respectively. Concurrently, comparing the results of the original and the rotated frames in Fig.~\ref{fig:8}, we see that $|\psi'_1(z, t)|^2$ provides the dynamics of the two inner solitons [see Fig.~\ref{fig:8}(d)], and $|\psi'_{-1}(z, t)|^2$ gives that of the two outer solitons [see Fig.~\ref{fig:8}(e)]. They both match with the dynamics of scalar solitons shown in Fig.~\ref{fig:1}(a) and Fig.~\ref{fig:1}(c), respectively for the inner and outer solitons. In general, depending on the extend of the initial overlap, the inner and outer solitons have different masses. The ratio of masses between the inner and outer solitons can be easily obtained via the wavefunctions in the rotated frame as, $$\frac{\int_{-\infty}^{\infty}|\psi_{1}^\prime|^2 dz}{\int_{-\infty}^{\infty}|\psi_{-1}^\prime|^2 dz}=\frac{1+k\Delta\csch{k\Delta}}{1-k\Delta\csch{k\Delta}}.$$ Hence, we can control the size of the inner and outer ferromagnetic solitons by varying $\Delta$. 

\begin{figure}
\centering
\includegraphics[width= \columnwidth]{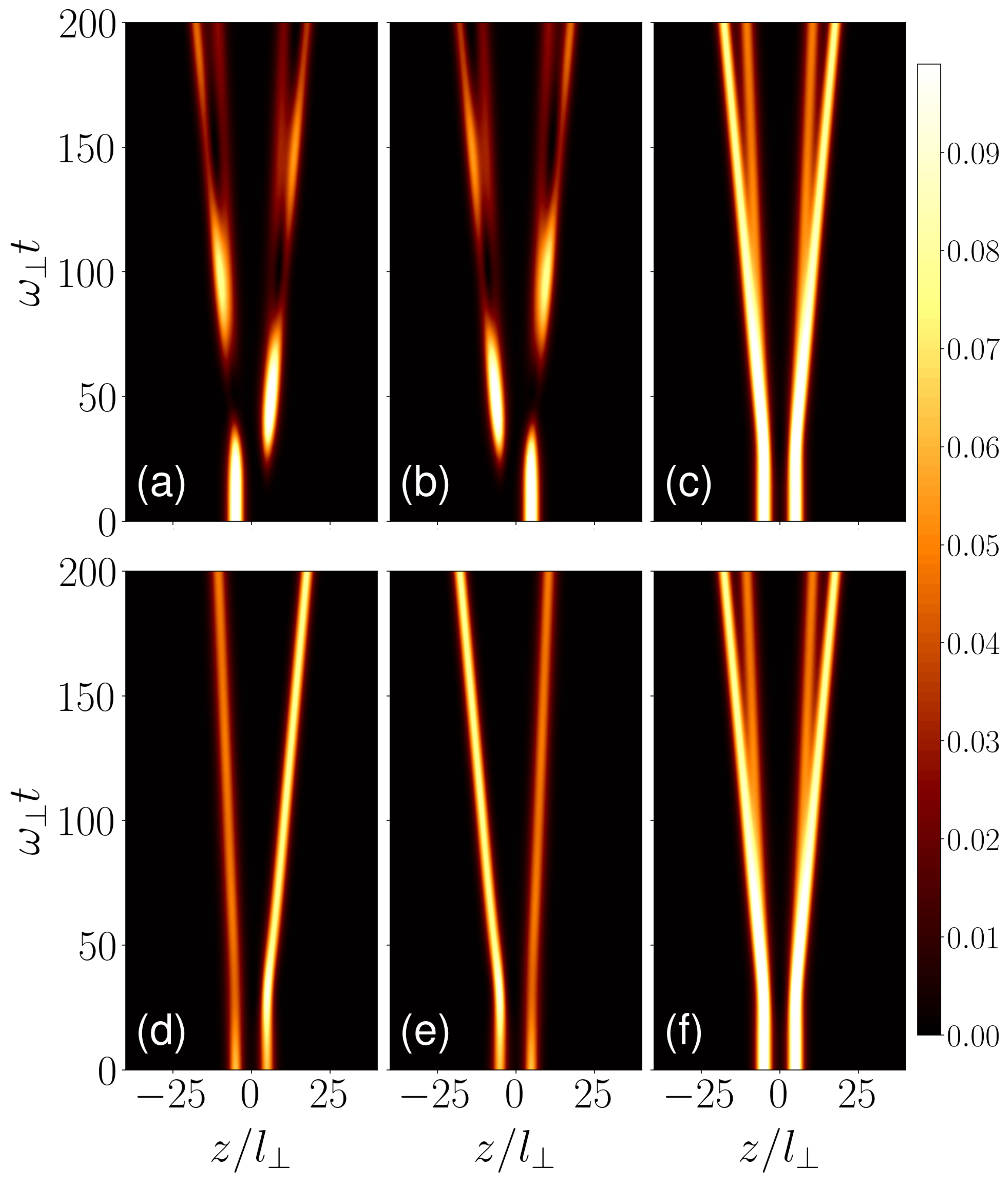}
\caption{\small{Dynamics of two overlapping polar bright solitons for $\gamma=1$, $\Delta=10l_\perp$, $\bar c_0/\hbar\omega_\perp=-2$, and $\phi_1=\pi/2$ rad. (a)-(c) show the dynamics of $|\psi_1|^2+|\psi_{-1}|^2$, $|\psi_0|^2$ and the total density $n(z, t)$, respectively. (d)-(f) show the dynamics of $|\psi'_1|^2$, $|\psi'_{-1}|^2$ and the total density $n'(z, t)=|\psi'_1|^2+|\psi'_{-1}|^2$, respectively. Note that $n(z,t)=n'(z,t)$.}}
\label{fig:10} 
\end{figure}

The population dynamics for $\phi_1=\pi$ rad is identical to that of $\phi_1=0$ as discussed above, but the spin density of inner and outer solitons are now flipped, compare Figs.~\ref{fig:9}(a) and \ref{fig:9}(c). For $\phi_1=\pi/2$ rad, population and spin density dynamics are qualitatively different from the $\phi_1=0$ case. The population dynamics for $\phi_1=\pi/2$ rad is shown in Fig.~\ref{fig:10} and the corresponding spin density dynamics is shown in Fig.~\ref{fig:9}(b). After the initial spin-mixing, the dynamics results in four ferromagnetic solitons and they move away from the centre on either side. Due to the initial phase difference of $\pi/2$, it takes a finite time to reach a miscible state as seen in Figs.~\ref{fig:10}(a)-\ref{fig:10}(c), via a transient oscillatory dynamics. The initial states in the rotated frame for $\phi_1=\pi/2$ are
\begin{eqnarray}
\label{ree01}
\psi_{\pm 1}'(z)&=&\frac{\sqrt{k}}{2}\left[\frac{\sech\left(k\left(z+\Delta/2\right)\right)}{\sqrt{2}}\pm \frac{i\sech\left(k\left(z-\Delta/2\right)\right)}{\sqrt{2}}\right]
\end{eqnarray}
representing two independent two soliton solutions, but with opposite phase gradients. As we know from the scalar case for $\phi=\pi/2$ rad [see Figs.~\ref{fig:1}(b) and \ref{fig:1}(e)], there is a transient atomic current along the direction of phase gradient. Therefore, in $\psi_1'(z)$, there is a transient atomic current from left to right, and in $\psi_{-1}'(z)$, it is from right to left. That results in a pair of asymmetric solitons in both $\psi_1'(z)$ and $\psi_{-1}'(z)$. Since the denser solitons move faster, the inner solitons are lighter, contrasting the $\phi_1=0$ case where all solitons are identical. As seen in Fig.~\ref{fig:9}(b), the spin vectors point in opposite directions among the inner solitons and the same among the outer ones. Again, by tuning $\Delta$, we can control the mass ratio between the inner and outer solitons, also the frequency of transient oscillation seen in the initial stage of the dynamics in Figs.~\ref{fig:10}(a) and \ref{fig:10}(b).
 
 \subsubsection{$\gamma=0.5$}
 \label{gh}
 
\begin{figure}
\centering
\includegraphics[width= \columnwidth]{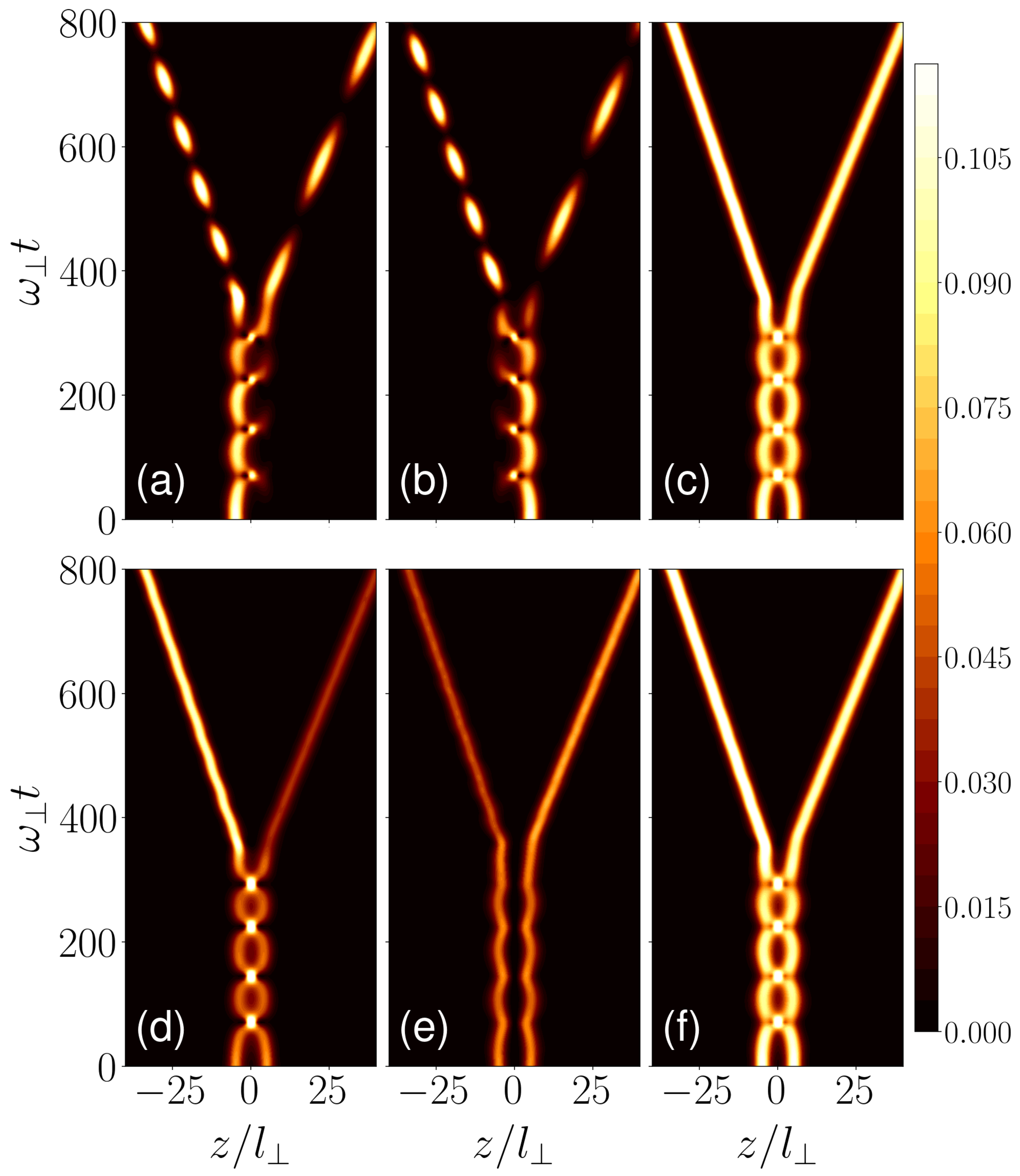}
\caption{\small{Dynamics of two overlapping polar bright solitons for $\gamma=0.5$, $\Delta=10l_\perp$, $\bar c_0/\hbar\omega_\perp=-2$, and $\phi_1=0$. (a)-(c) show the dynamics of $|\psi_1|^2+|\psi_{-1}|^2$, $|\psi_0|^2$ and the total density $n(z, t)$, respectively. (d)-(f) show the dynamics of $|\psi'_1|^2$, $|\psi'_{-1}|^2$ and the total density $n'(z, t)=|\psi'_1|^2+|\psi'_{-1}|^2$, respectively. Note that $n(z,t)=n'(z,t)$.}}
\label{fig:11} 
\end{figure}

\begin{figure}
\centering
\includegraphics[width= \columnwidth]{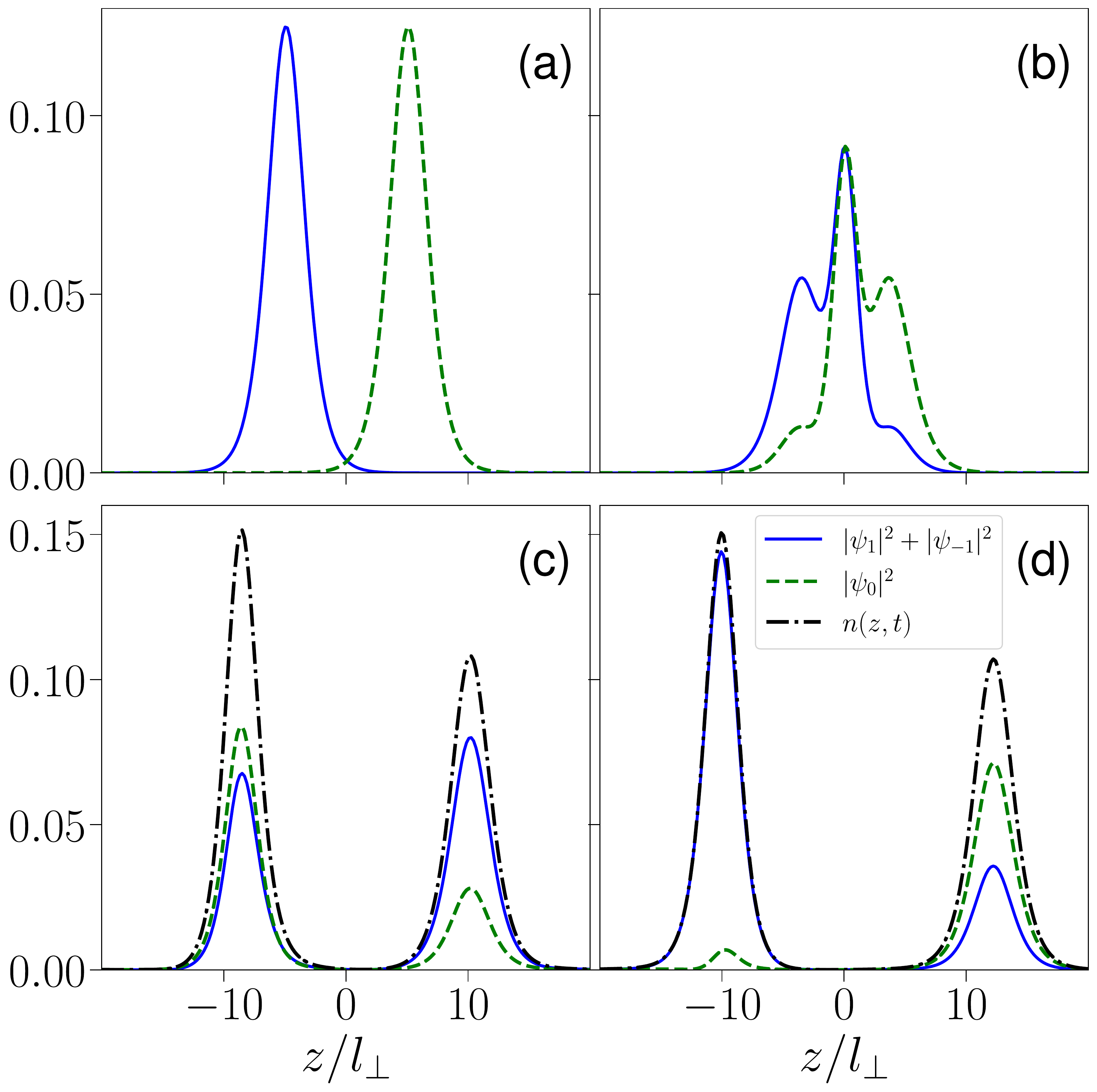}
\caption{\small{(a) The snapshot of the initial state of two polar solitons. (b) The snapshot of densities at $\omega_\perp t=220$ (c) The density profile of the left oscillaton at $\omega_\perp t=420$. (d) The density profile of the right oscillaton at $\omega_\perp t=445$. For all plots $\gamma=0.5$, $\Delta=10l_\perp$, $\bar c_0/\hbar\omega_\perp=-2$, and $\phi_1=0$. The solid lines shows $|\psi_1|^2+|\psi_{-1}|^2$}, dashed lines show $|\psi_0|^2$ and dotted-dashed lines show the total density $n(z)$.}
\label{fig:12} 
\end{figure}

In the following, we analyze the dynamics for $\gamma=0.5$ and observe the emergence of a pair of non-identical oscillatons propagating in opposite directions. Oscillatons are solitons in which the total density profile remains stationary while the populations in different spin components oscillate with a constant frequency \cite{sza11,sza10}. In Figs.~\ref{fig:11}(a)-\ref{fig:11}(c), we show the dynamics of $|\psi_1|^2+|\psi_{-1}|^2$, $|\psi_0|^2$ and the total density $n(z, t)$, respectively for $\gamma=0.5$, $\Delta=10 l_\perp$ and $\phi_1=0$. In the initial stage, spin-mixing leads to oscillatory dynamics. Later, the condensates transform into a pair of independent oscillatons moving in opposite directions. Unlike the case of $\gamma=1$ where we see four final solitons, for $\gamma=0.5$, we see only two final solitons. It is explained later using the wavefunctions in the new frame. The initial oscillatory dynamics takes place between the two configurations shown in Figs.~\ref{fig:12}(a) and \ref{fig:12}(b). Figure~\ref{fig:12}(b) is a completely miscible state with peaks of $|\psi_{\pm 1}|^2$ coincide with that of $|\psi_{0}|^2$, which implies that the spin-mixing is taking place. On the oscillatory dynamics, each spin component leaves a small trail of atoms on the opposite sides, i.e., $m=\pm 1$ in the right region and $m=0$ in the left region. Eventually, they cause the formation of a pair of oscillatons. Figures~\ref{fig:12}(c) and \ref{fig:12}(d) are density snapshots taken at two different instants after the oscillatons are formed. The total density $n(z, t)$ of each oscillaton is identical in both figures, whereas the densities of different components vary in time due to the spin-mixing. The total population oscillates between $m=0$ and $m=\pm 1$ within each oscillaton. In general, the frequency of internal oscillations of the two oscillatons are different and can be tuned via the extent of overlap, i.e. varying $\Delta$. Since $\langle p\rangle=0$, the denser oscillaton travels slower than the other.

Before the oscillatons are formed, the whole system reaches a miscible state. The density pattern of this miscible state is susceptible to the initial noise in the system, if any present. The latter can thereby affect the properties of the final oscillatons, such as the mass asymmetry,  velocities and the frequency of the internal oscillations. But general qualitative features of the dynamics remain the same. Such sensitivity to initial noise is not seen in any other cases discussed here.

\begin{figure}
\centering
\includegraphics[width= \columnwidth]{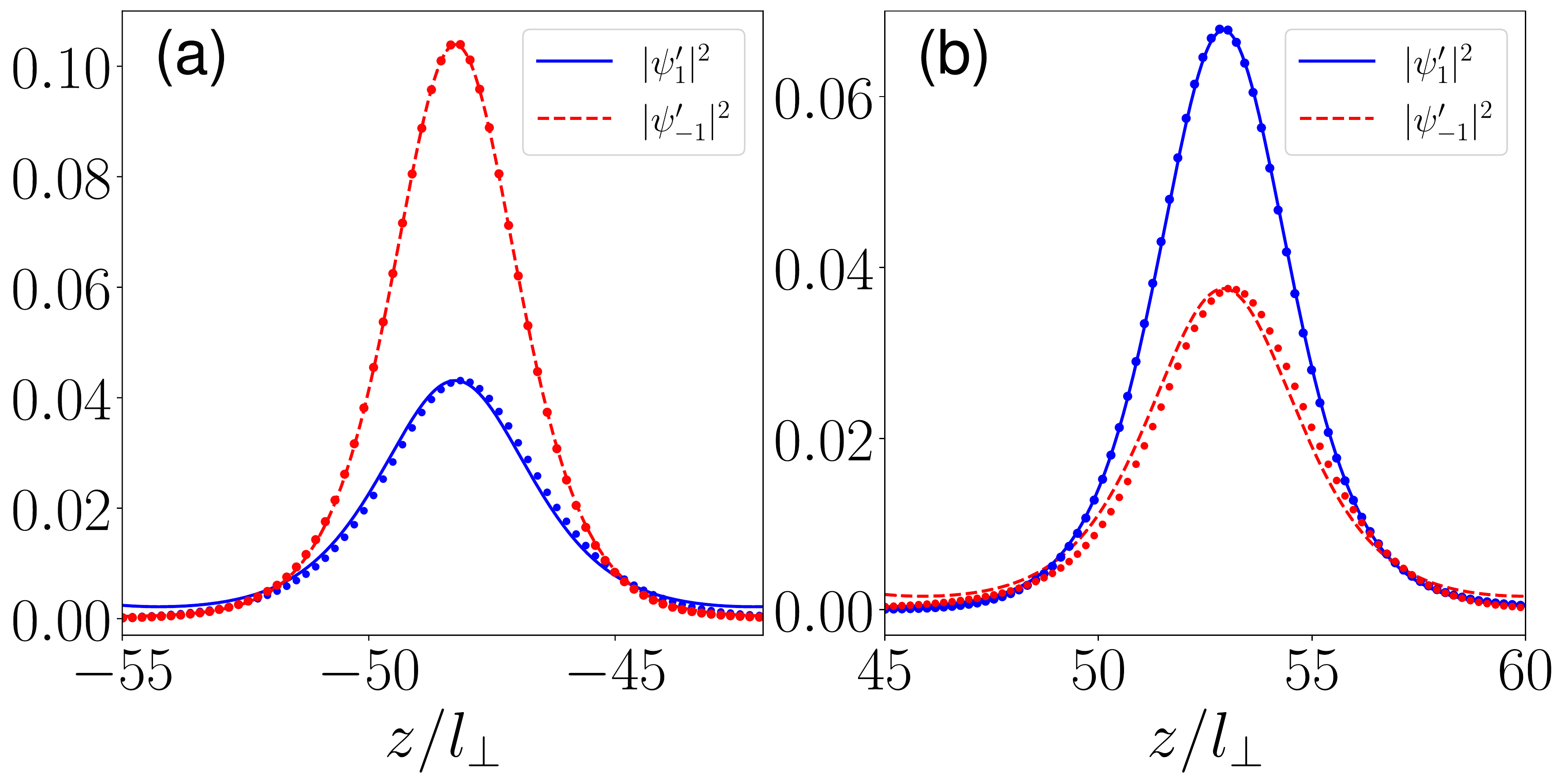}
\caption{\small{Density profiles of the oscillaton in the rotated frame at two different instants. The solid ($|\psi_1'(z)|^2$) and dashed ($|\psi_{-1}'(z)|^2$) lines show the the stationary solution obtained by solving Eqs.~(\ref{oes}). The dotted line shows the solutions from the dynamics. (a) for the oscillaton in the left and (b) for the oscillaton in the right. Both snap shots are taken at $\omega_\perp t=1000$.}}
\label{fig:13} 
\end{figure}

For $\phi_1=0$, the initial states in the rotated frame are given by the two soliton solutions in Eq.~(\ref{re01}). For $\gamma=0.5$, in the rotated frame, we have a binary condensate with attractive intra and inter-component interactions governed by the Eq.~(\ref{efe}). Therefore both $\psi_1'(z)$ and $\psi_{-1}'(z)$ have the tendency to stick together in the same spatial regions, and leading to two final oscillatons or solitons. Note that, unlike in the lab frame, an oscillaton is characterized by stationary density profile for each component in the rotated frame, as seen in Figs.~\ref{fig:11}(d)-\ref{fig:11}(e). The wavefunctions for a stationary oscillation in the new frame can be written as $\psi'_{\pm 1}(z)=\eta_{\pm}\exp[i(\mu_{\pm}t+\phi_{\pm})]$ \cite{sza11,sza10}, where $\eta_{\pm}$ satisfy two coupled ordinary differential equations, 
\begin{equation}
\left(-\frac{\hbar^2}{2m}\frac{d^2}{dz^2}+(\bar c_0+\bar c_1)\eta_\pm^2+(\bar c_0-\bar c_1)\eta_\mp^2\right)\eta_\pm=\mu_\pm\eta_\pm.
\label{oes}
\end{equation}
Taking the values of $\mu_\pm$ and $\phi_\pm$ from the numerical results and solving Eqs.~(\ref{oes}) we obtain the stationary solutions $\eta_\pm$. The excellent agreement shown in Fig.~\ref{fig:13} confirm us that the final solitons formed in the dynamics are oscillatons. In Figs.~\ref{fig:13}(a) and \ref{fig:13}(b), we show the density profiles of $\psi'_{\pm 1}(z)$ for the oscillaton from numerics (dotted lines) and the solution of Eqs.~(\ref{oes}) (solid and dashed lines) at two different instants. Figure~\ref{fig:13}(a) is for the left and Fig.~\ref{fig:13}(b) is for the right oscillaton.

\begin{figure}
\centering
\includegraphics[width= \columnwidth]{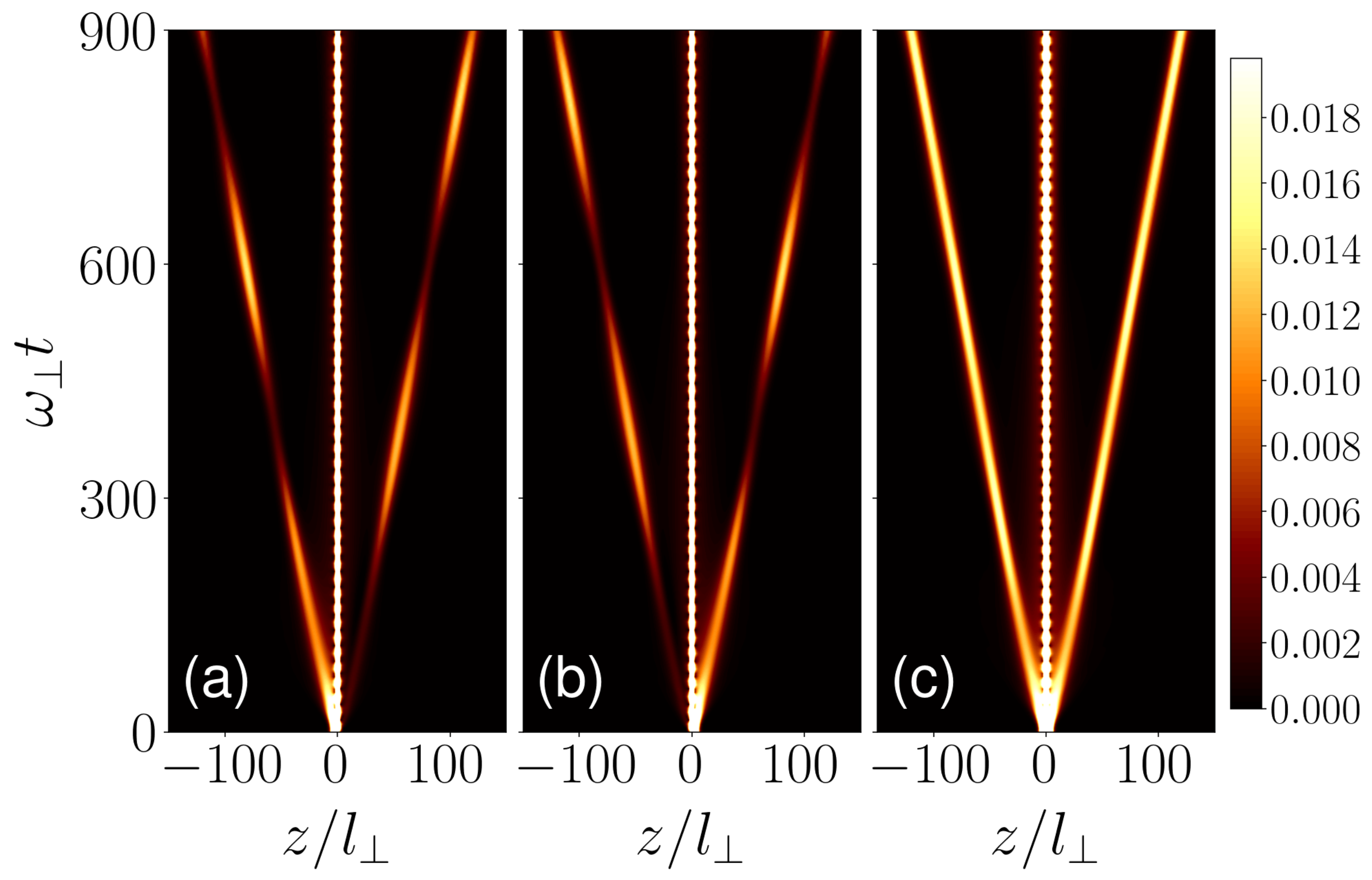}
\caption{\small{Dynamics of two overlapping polar bright solitons for $\gamma=0.5$, $\Delta=5l_\perp$, $\bar c_0/\hbar\omega_\perp=-2$, and $\phi_1=0$ rad. (a) show the dynamics of $|\psi_1|^2+|\psi_{-1}|^2$, (b) of $|\psi_0|^2$ and (c) of the total density $n(z, t)$, respectively.}}
\label{fig:14} 
\end{figure}

Interestingly, for $\phi_1=0$, increasing the extent of overlap or decreasing $\Delta$ introduces a qualitatively new feature to the dynamics. Contrary to $\Delta=10 l_\perp$ case, an additional stationary ferromagnetic soliton is formed, centerd at $z=0$, see Fig.~\ref{fig:14}. Comparing Figs.~\ref{fig:14} and \ref{fig:11}, we see that decreasing $\Delta$ reduces the frequency of the internal oscillations and increases the velocity of the final oscillatons. The ferromagnetic soliton exhibits a strong breathing character due to its dynamical formation.

\begin{figure}
\centering
\includegraphics[width= \columnwidth]{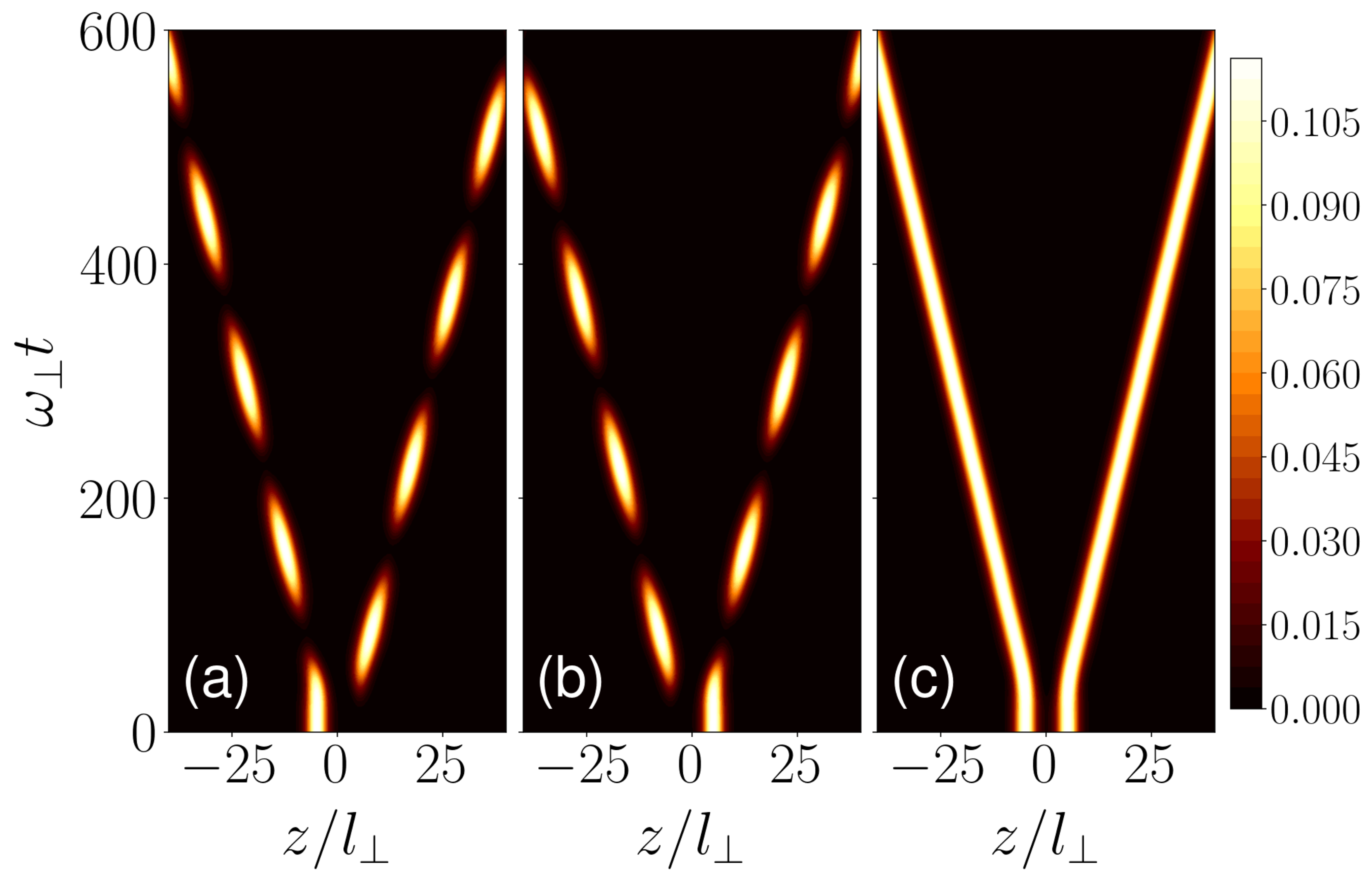}
\caption{\small{Dynamics of two overlapping polar bright solitons for $\gamma=0.5$, $\Delta=10l_\perp$, $\bar c_0/\hbar\omega_\perp=-2$, and $\phi_1=\pi/2$ rad. (a) show the dynamics of $|\psi_1|^2+|\psi_{-1}|^2$, (b) of $|\psi_0|^2$ and (c) of the total density $n(z, t)$, respectively.}}
\label{fig:15} 
\end{figure}

Finally, we discuss the case of $\phi_1=\pi/2$ rad. Unlike that of $\phi_1=0$, the two final oscillatons are symmetric [see Fig.~\ref{fig:15}]. There is no transient oscillatory dynamics at the initial stage of the dynamics because of repulsion emerging from the initial phase difference of $\pi/2$. Also, spin mixing happens independently in the left and right regions due to the initial repulsion. Consequently, we have independent and identical oscillatons moving away from each other. Also, the noise does not affect the dynamics in contrast to the case of $\phi_1=0$.

 \section{Summary and Outlook}
 \label{sum}
In summarizing, we have analyzed the dynamics of two overlapping polar bright solitons in spin-1 condensates, which depends critically on the relative phase, the extent of overlap and the ratio between spin-independent and spin-dependent interactions. The same dynamics of scalar solitons revealed interesting scenarios, particularly the possibility of observing atomic switching. Atomic switching can find applications in implementing atom-based networks, identical to optical networks. Overlapping polar solitons resulted in non-trivial dynamics in spatial and spin degrees of freedom. For vanishing spin-dependent interactions, we observed Josephson like oscillations of each component in an effective double-well potential created by the density of other components. For identical spin-dependent and spin-independent interactions, we observed the formation of four ferromagnetic solitons. In the last case, when the spin-dependent interaction strength is half of the spin-independent one, the dynamics led to the formation of two oscillatons. Strikingly, the properties of the final bright solitons can be easily tuned by the initial state and the interaction parameters. Our studies offer a new possibility for engineering matter waves. Extending the above analysis to a pair of overlapping ferromagnetic solitons and ferromagnetic-polar solitons may lead to exciting and completely new dynamics than those reported here. We also expect complex dynamics to emerge from overlapping more than two solitons.
 
  \section{Acknowledgments}
  \label{ack}
 R.N. acknowledges support from DST-SERB for the Swarnajayanti fellowship File No.~SB/SJF/2020-21/19. We acknowledge National Supercomputing Mission (NSM) for providing computing resources of 'PARAM Brahma' at IISER Pune, which is implemented by C-DAC and supported by the Ministry of Electronics and Information Technology (MeitY) and Department of Science and Technology (DST), Government of India. G. H. acknowledges the funding from DST India through an INSPIRE scholarship.
\bibliographystyle{apsrev4-1}
\bibliography{lib.bib}
\end{document}